\documentclass[10pt,conference]{IEEEtran}

\newcommand{\hpcayear}{2027}

\usepackage{amsmath,amssymb,amsfonts}
\usepackage{mathtools}
\usepackage{amsthm}
\newtheorem{definition}{Definition}

\usepackage{graphicx}
\usepackage{xcolor}
\usepackage{fancyhdr}
\usepackage{subcaption}
\usepackage{multirow}
\usepackage{array}
\usepackage{booktabs}

\usepackage{algorithm}
\usepackage{algpseudocode}

\usepackage{enumitem}

\usepackage[T1]{fontenc}
\usepackage{pifont}
\usepackage[utf8]{inputenc}
\DeclareUnicodeCharacter{2717}{\ding{55}} 

\usepackage[hyphens]{url}
\usepackage{cite}
\usepackage[hidelinks]{hyperref}
\usepackage{cleveref}

\def\BibTeX{{\rm B\kern-.05em{\sc i\kern-.025em b}\kern-.08em
    T\kern-.1667em\lower.7ex\hbox{E}\kern-.125emX}}

\newcommand{\myparagraph}[1]{\noindent\textbf{#1}}

\newcommand{\papercount}{12}

\newcommand{\csthreecount}{6}
\newcommand{\cstwopaper}{97}

\newcommand{\myclearpage}[0]{\clearpage\newpage\setcounter{page}{1}}
\renewcommand{\myclearpage}[0]{}

\title{Rosetta: Automating First-Principles Performance Modeling Using Multi-Agent LLMs}

\newcommand\hpcaauthors{Anonymous Author(s)}
\newcommand\hpcaaffiliation{Anonymous Institution}
\newcommand\hpcaemail{anon@example.com}

\author{
  \ifdefined\hpcacameraready
    \IEEEauthorblockN{\hpcaauthors{}}
      \IEEEauthorblockA{
        \hpcaaffiliation{} \\
        \hpcaemail{}
      }
  \else
    \IEEEauthorblockN{Karthikeyan Sankaralingam, NVIDIA
    }
  \fi 
}

\fancypagestyle{camerareadyfirstpage}{%
  \fancyhead{}
  
  \fancyhead[C]{
    \ifdefined\aeopen
    \parbox[][12mm][t]{13.5cm}{\hpcayear{} IEEE International Symposium on High-Performance Computer Architecture (HPCA)}    
    \else
      \ifdefined\aereviewed
      \parbox[][12mm][t]{13.5cm}{\hpcayear{} IEEE International Symposium on High-Performance Computer Architecture (HPCA)}
      \else
      \ifdefined\aereproduced
      \parbox[][12mm][t]{13.5cm}{\hpcayear{} IEEE International Symposium on High-Performance Computer Architecture (HPCA)}
      \else
      \parbox[][0mm][t]{13.5cm}{\hpcayear{} IEEE International Symposium on High-Performance Computer Architecture (HPCA)}
    \fi 
    \fi 
    \fi 
    \ifdefined\aeopen 
      \includegraphics[width=12mm,height=12mm]{ae-badges/open-research-objects.pdf}
    \fi 
    \ifdefined\aereviewed
      \includegraphics[width=12mm,height=12mm]{ae-badges/research-objects-reviewed.pdf}
    \fi 
    \ifdefined\aereproduced
      \includegraphics[width=12mm,height=12mm]{ae-badges/results-reproduced.pdf}
    \fi
  }
  \fancyfoot[C]{}
}
\begin{document}
\maketitle

\ifdefined\hpcacameraready 
  \thispagestyle{camerareadyfirstpage}
  \pagestyle{empty}
\else
  \thispagestyle{plain}
  \pagestyle{plain}
\fi

\newcommand{\hpcaheight}{0mm}
\ifdefined\eaopen
\renewcommand{\hpcaheight}{12mm}
\fi


\begin{abstract}
Analytical performance models --- derivations of throughput or speedup
from hardware parameters --- make claims independently verifiable and
expose binding constraints, yet rarely accompany architecture papers
because building one by hand takes weeks of expert effort. We present
Rosetta, a multi-agent LLM pipeline that automatically
generates first-principles analytical models from research paper PDFs.
Given a paper as sole input, Rosetta produces a mathematical
specification, an executable Python model, and a plain-English
interpretation --- all autonomously, with zero human intervention.
The formalization process itself is the primary value: it surfaces
implicit assumptions and identifies missing parameters.
Four design decisions address failure modes of na\"ive LLM-based
generation: a scientific constitution that prohibits circular reasoning,
verify-repair loops with independent critic agents, dual verification
separating functional correctness from scientific validity, and a
best-of-$N$ ensemble that exploits LLM stochasticity.

We evaluate Rosetta across three complementary tracks: expert
evaluation of \papercount{} landmark papers (CS1), automated scoring
of \cstwopaper{} unfiltered ISCA~2025 and HPCA~2026 papers (CS2), and
author self-evaluation by six active research groups (CS3).
Across CS1, specification quality scores 4--5/5 on 10 of 12 papers
with zero significant hallucinations; across CS2, 56\% of fit-screened
papers reach Tier~A insight quality.
The strongest finding comes from CS3: Rosetta's output led to revised
claims and new experiments in active submissions, and five of six
author-evaluators said they would use it again.
\end{abstract}


\section{Introduction}
\label{sec:introduction}

An analytical performance model — a closed-form or small-program derivation that predicts a system's throughput, latency, energy, or speedup from hardware parameters and workload characteristics — is one of the most useful artifacts a computer architecture paper can provide. It makes transparent questions that cycle-level simulators can also address but require significant instrumentation and infrastructure to answer: \emph{why} performance bounds exist, which parameters are the binding constraints, and how results would change under different conditions. Such models are invaluable across the research lifecycle. They serve readers by verifying claims independently, help authors debug designs alongside simulation, provide self-contained reproducible artifacts, allow industry practitioners to share architectural insights without exposing proprietary simulation infrastructure or IP (e.g., Anton~\cite{TODO-anton}, a  molecular dynamics ASIC evaluated in our case-study), and serve as artifacts that guide follow-on work.

Despite this value, rigorous analytical models rarely accompany architecture papers. Building one by hand requires weeks of skilled effort, mathematical fluency, workload and hardware intuition. Consequently, the research community recognizes the value of analytical models but overwhelmingly relies on simulation or measurement results instead.

\begin{figure}[t]
  \centering
  \includegraphics[width=\columnwidth]{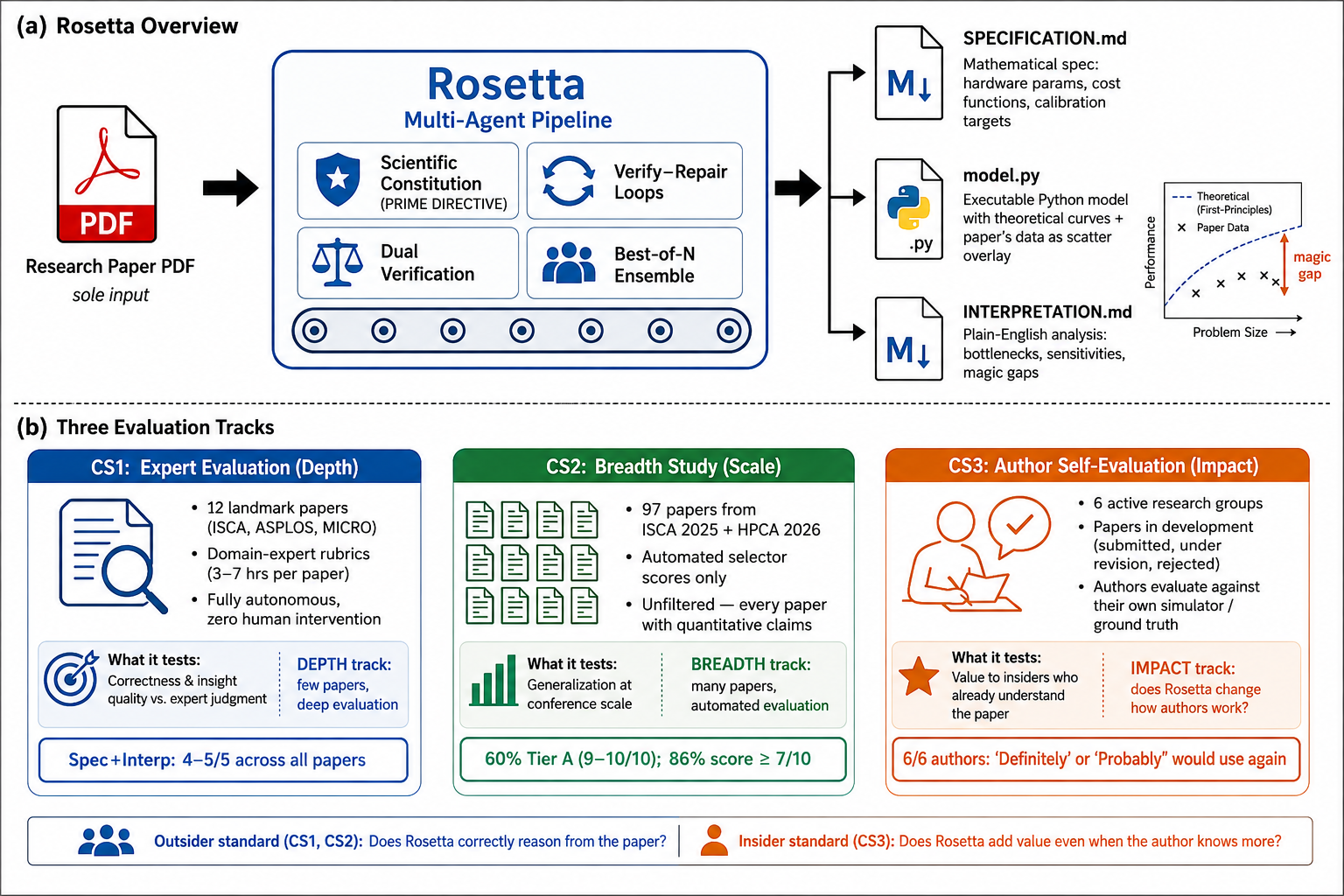}
  \caption{Rosetta and Paper Overview}
  \vspace{-0.25in}
  \label{fig:paper-overview}
\end{figure}

We present \textbf{Rosetta}, a closed-loop multi-agent pipeline that \textbf{automates} first-principles analytical model generation from research paper PDFs (Figure~\ref{fig:paper-overview}). Given a PDF as its sole input, Rosetta produces three artifacts: (i)~\texttt{SPEC.md}, a mathematical specification of hardware parameters and cost functions; (ii)~\texttt{model.py}, a self-contained executable Python model; and (iii) \texttt{INTERPRETATION.md}, a plain-English explanation of the model's insights. A central concept is to quantify any discrepancy between what the paper's described mechanism can achieve from first principles and its reported results, highlighting either unstated assumptions, omitted optimizations, or conservative claims. Just as code and data releases became standard practice, Rosetta-generated artifacts can make performance claims independently verifiable by any reader, without access to the authors' simulation infrastructure.

We evaluate Rosetta in fully autonomous mode on published research papers—a deliberate stress test. A common misconception is that a camera-ready PDF is a perfectly precise and complete artifact, rendering automated modeling redundant. In reality, published papers are highly compressed and routinely elide implicit assumptions or obscure hardware parameters. By succeeding on these sparse documents, we establish a lower bound on Rosetta's utility. If it can derive an independent, verifiable artifact from a compressed PDF to serve readers and reviewers, it is even more capable when applied to the ongoing, parameter-rich drafts of active researchers (as demonstrated in Case Study 3) or the highly detailed technical specifications used in industry.
More broadly, the core question Rosetta answers for any technical document is: \emph{did I build what I think I built, and are the benefits and limitations what I claimed they were?}  Automating this process raises four challenges: LLMs defaulting to circular reasoning (using reported results as inputs rather than calibrating against them), low-quality single-shot specifications, code that faithfully implements scientifically invalid specs, and LLM stochasticity. Rosetta addresses these via a \textbf{scientific constitution} (the PRIME DIRECTIVE) enforcing first-principles derivation, a \textbf{verify-repair loop} with independent critic agents, \textbf{dual verification} separating functional correctness from scientific validity, and a \textbf{best-of-$N$ ensemble} to ensure output quality and insight.

Recent work has begun to argue for GenAI-assisted discovery in architecture more broadly~\cite{sankaralingam2026computerarchitecturesalphazeromoment,gupta2026archagentagenticaidrivencomputer}; Rosetta addresses one specific, thus far unresolved tool problem in that emerging space.

\subsection*{Contributions}
\begin{itemize}[noitemsep,leftmargin=*]

\item \textbf{Rosetta}: an end-to-end multi-agent pipeline that
  produces first-principles analytical performance models from research
  paper PDFs, with a median end-to-end wall-clock time of 72~minutes,
  reducible to $\sim$25~minutes by running ensemble instances in parallel.
  It is released open-source with this paper,
    including all code, prompts, and evaluation artifacts, with
    Anthropic API implementation.

\item \textbf{The PRIME DIRECTIVE}: a shared scientific constitution
  enforcing first-principles derivation, calibration-as-overlay,
  self-containment, and baseline vs.\ proposed comparison across all
  agents. We show empirically that removing the constitution degrades model
  quality substantially.

\item \textbf{A verify-repair architecture with dual verification}:
  independently separating functional correctness (spec-to-code
  fidelity) from scientific validity (constitution compliance), closing
  a failure mode where a faithful implementation of a bad spec passes
  single-verifier review.

\item \textbf{Evaluation across three case study (CS) tracks}:
  CS1 evaluates \papercount{} landmark architecture papers against
  domain-expert rubrics; CS2 applies Rosetta in breadth to \cstwopaper{} ISCA~2025 and HPCA~2026
  papers without expert evaluation; CS3 gives Rosetta to \csthreecount{} researchers and
  collects author self-evaluations.

\item \textbf{CAAM-Bench}: the first benchmark for evaluating analytical
  model generators on architecture papers --- \papercount{} landmark papers
  with human-vetted reference artifacts (\texttt{SPECIFICATION.md} and
  \texttt{model.py}) and a four-dimension scoring rubric.
  Rosetta's CS1 scores serve as the published baseline for future tools.
\end{itemize}

\subsection*{Key Findings}
\begin{itemize}[noitemsep,leftmargin=*]

\item \textbf{The formalization itself is the primary value.}
Every CS3 author named \texttt{SPECIFICATION.md} — not the quantitative
model — as the most valuable output. The act of reverse-engineering a paper's performance structure into a
mathematical specification surfaces implicit assumptions, flags missing
parameters, and produces a readable analytical skeleton — value that is
decoupled from whether the model achieves numerical accuracy. In two CS3 cases, Rosetta's formalization output directly caused
changes to the paper: a simulation methodology comparison in one case,
and a sharpened separation of GPU and custom-accelerator contributions
in the other.

\item \textbf{Human evaluators confirm correctness and insight.}
Across \papercount{} CS1 papers, independent domain-competent researchers
scored specification quality at 4--5/5 on 10 of 12 papers with zero
significant hallucinations, and rated 11 of 12 papers at 4/5 or 5/5
overall usefulness. No evaluator said ``No'' to future use.

\item \textbf{Rosetta generalizes at scale.}
In CS2, 60\% of \cstwopaper{} unfiltered conference papers score in
Tier~A (9--10/10) on the automated selector's insight quality metric,
and 86\% score 7/10 or above.

\item \textbf{Researchers want to use it again --- and it changed their papers.}
Five of six CS3 evaluators said ``Definitely'' for future use; four had
concrete paper impact (revised claims, new experiments, sharpened exposition).
The adoption signal is not about numerical accuracy: as one evaluator put it,
\textit{``No reviewer would have the time to write a simulation to validate
a paper's empirical results, but with Rosetta it's a 5-minute task.''}
Another noted it \textit{``may have helped us earlier for idea development
and eliminating less promising ideas.''}
\end{itemize}

\myparagraph{Paper organization.}
Section~\ref{sec:formulation} provides Rosetta's problem definition and design goals.
Section~\ref{sec:design} describes the system design, including empirical
evidence for each design decision (Section~\ref{sec:why_naive_fails}).
Section~\ref{sec:eval-methodology} frames the evaluation methodology.
Sections~\ref{sec:cs1}--\ref{sec:cs3} present the three case study
tracks. Section~\ref{sec:related_work} surveys related work.
Section~\ref{sec:conclusion} concludes.
\myclearpage
\section{The Problem Formulation}
\label{sec:formulation}

The standard artifact of architecture evaluation is a simulator or measurement
harness --- a tool that produces performance numbers but embeds its assumptions
in code rather than exposing them mathematically.
When a paper's performance claims rest on such tools, independent verification
requires access to the full infrastructure.
Rosetta's task is to produce the complementary artifact: a mathematical
specification and executable model derived solely from the paper's prose,
against which the paper's claims can be independently checked.
We define this task as follows.

\begin{definition}[Analytical Model Generation]
Given a research paper PDF $P$ describing a computer architecture or
systems contribution with performance claims, produce: (i)~a
mathematical specification $\mathcal{S}$ defining the hardware
parameters, workload model, and analytical cost functions for both the
baseline and proposed systems, derived from the paper's prose without
using reported results as inputs; (ii)~an executable model $M$
implementing $\mathcal{S}$; and (iii)~an interpretation $\mathcal{I}$
reporting the model's findings, including any quantified discrepancy
between $M$'s first-principles prediction and the paper's reported
results.
\end{definition}

The key constraint is \emph{first-principles derivation}: $M$ must
compute performance from physical parameters ($\mathcal{S}$), not from
the paper's reported results.  The paper's results appear in $M$'s
output as calibration scatter points --- a visual representation of
whether the first-principles prediction and the reported result agree.
When they disagree significantly, the discrepancy is the \emph{magic
gap}.  This formulation reflects Box's principle that ``all models are
wrong, but some are useful''~\cite{box1976science}: the goal is not
numerical precision but mechanistic understanding that makes the paper's
performance claims independently auditable.

This formulation has a natural failure mode: if the paper does not
expose the hardware parameters needed to derive performance from first
principles, $M$ cannot be fully derived.  It can still produce bounds,
calibration-mode estimates labeled as such, and a diagnostic listing of
which parameters are missing.  This is the correct behavior when the
paper's study reports measured speedups without
exposing the underlying hardware configuration.


\section{System Design}
\label{sec:design}

\begin{figure}[t]
  \centering
  \includegraphics[width=\columnwidth]{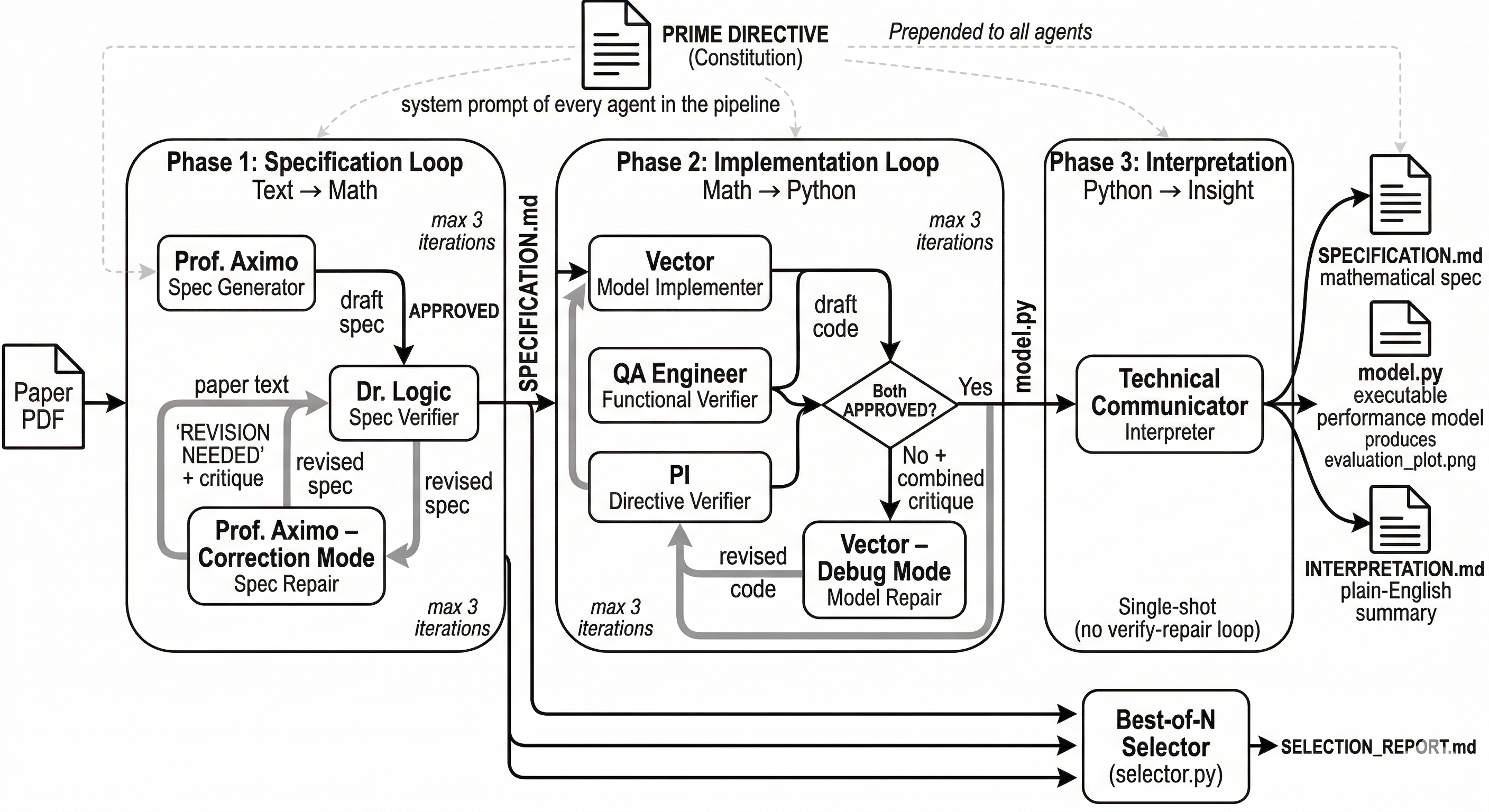}
  \caption{Rosetta pipeline. Three sequential phases transform a paper PDF into
    a mathematical spec, executable model, and interpretation. Phases 1 and 2 run
    verify-repair loops of up to three iterations. The full pipeline runs $N{=}3$ times;
    a selector agent picks the best run.}
  \label{fig:system-overview}
\end{figure}

Rosetta is a closed-loop pipeline that transforms a research paper PDF into three
artifacts: a mathematical specification (\texttt{SPECIFICATION.md}), an executable
Python performance model (\texttt{model.py}), and a plain-English interpretation
(\texttt{INTERPRETATION.md}).
Figure~\ref{fig:system-overview} shows the overall architecture.
The pipeline is organized into three sequential phases, each driven by one or more
large language model (LLM) agents operating under a shared scientific constitution.
To mitigate LLM stochasticity, the entire pipeline runs multiple times and a
selector agent chooses the best output (Section~\ref{sec:ensemble}).
The complete system --- including all agent prompts, the PRIME DIRECTIVE
constitution, and evaluation artifacts --- is released as open-source
with this paper; the prompts are the primary design artifact and are
integral to reproducibility. A concise usage guide accompanies the release to help users interpret outputs
without reading this paper or becoming LLM experts.

\subsection{Phase 1: Specification (Text $\rightarrow$ Math)}
\label{sec:phase1}

The goal of Phase 1 is to reverse-engineer a rigorous mathematical specification
from the paper's prose — the ``physics'' of the described system expressed as
concrete variables, cost functions, and workload models.

\myparagraph{Generation (Prof.\ Aximo).}
The specification generator, characterized as a rigorous expert in analytical
performance modeling who ``only trusts variables, equations, and concrete data
points,'' reads the full paper text and produces a structured document containing:
(i)~all hardware parameters with Python-compatible names and suggested value ranges
(e.g., \texttt{dram\_bandwidth\_gbs = 51.2});
(ii)~a workload model specifying the synthetic input distribution the model will use;
(iii)~analytical cost functions for both the \emph{baseline} and \emph{proposed}
systems, derived bottom-up from hardware constraints;
(iv)~calibration targets — concrete data points extracted from the paper's figures
and tables, formatted as \texttt{(input\_state, claimed\_output, source)} tuples;
and
(v)~an initial ``magic gap'' hypothesis noting any visible discrepancy between
what the mechanism implies and what the paper reports.

\myparagraph{Verification (Dr.\ Logic).}
An independent verifier — a ``pedantic senior technical reviewer'' — audits the
specification against the original paper and the shared constitution
(Section~\ref{sec:constitution}).
It checks for missing variables, hallucinated formulas, absent calibration data,
and — most critically — circular reasoning: using the paper's own reported results
as formula inputs rather than as calibration overlays.
The agent responds with exactly \texttt{APPROVED: <reason>} or
\texttt{REVISION NEEDED: <detailed critique>}.
The complete critique is saved to disk regardless of outcome, creating an audit trail.

\myparagraph{Repair loop and audit trail.}
If the spec is rejected, the generator agent is re-invoked in ``Correction Mode''
with the original paper, the rejected spec, and Dr.\ Logic's critique.
The loop runs for up to three iterations; in practice, most specifications
converge within one or two.
Every intermediate artifact is saved (draft specs, draft models, all critique
files), providing full transparency: a researcher can trace exactly how each
conclusion was reached, and a diagnosis tool evaluates whether each iteration
made genuine progress.

\subsection{Phase 2: Implementation (Math $\rightarrow$ Python)}
\label{sec:phase2}

Phase 2 translates the approved specification into a self-contained, executable
Python performance model.

\myparagraph{Implementation (Vector).}
The implementer agent — a ``brilliant Python data scientist and simulation
engineer'' — receives both the approved specification and the original paper text
and produces a Python script with a fixed structure:
(i)~a \texttt{Config} dataclass holding all parameters from the spec;
(ii)~\texttt{cost\_baseline()} and \texttt{cost\_proposed()} functions implementing
the spec's cost formulas exactly;
(iii)~a simulation loop sweeping over the parameter ranges defined in the spec;
and (iv)~a \texttt{matplotlib} visualization with a dashed blue baseline curve,
a solid green proposed curve, and red~\texttt{X} scatter markers for the paper's
claimed results, saved as \texttt{evaluation\_plot.png} (never \texttt{plt.show()},
since the pipeline runs headless).
Generated code is syntax-validated via Python's \texttt{ast.parse()} before saving.

\myparagraph{Dual verification.}
Phase 2 uses \emph{two independent verifiers} because code quality has two
orthogonal failure modes.
The \textbf{QA Engineer} performs a line-by-line functional audit: do variable
names match the spec, are formulas implemented correctly, will numpy broadcasting work?
The \textbf{PI (Principal Investigator)} performs a scientific validity audit: is
there a baseline vs.\ proposed comparison, are calibration overlays present, is the
code self-contained, and — most importantly — is performance derived from physical
constraints rather than from table data?
The PI's test is concrete: ``If you removed all Table/Figure data from the code,
would it still calculate performance? If not, it is violating first principles.''
Both agents must return \texttt{APPROVED} for the loop to exit; either rejection
triggers a repair.

\myparagraph{Repair loop.}
The implementer agent, in ``Debug Mode,'' receives both the QA Engineer's and PI's
critiques simultaneously and rewrites the model from scratch addressing all issues.
The loop again runs for up to three iterations, saving all draft versions and
critique files as an audit trail.

\subsection{Phase 3: Interpretation (Python $\rightarrow$ Insight)}
\label{sec:phase3}

The final phase is a single-shot generation by a Technical Communicator agent.
It reads the final \texttt{SPECIFICATION.md} and \texttt{model.py} and produces
a plain-English document covering: the workload the model assumes; how the baseline
is modeled and what it implies; the proposed mechanism and what changes; the
critical variables and threshold conditions where the proposed system wins or loses;
and any gaps between the theoretical curves and the paper's claimed results.
The tone guidance is ``insightful and explanatory — explain model semantics, not
Python syntax.''
This phase has no verify-repair loop: interpretation quality improves less from
iteration than specification or code correctness.

\subsection{The Scientific Constitution}
\label{sec:constitution}

Every agent in the pipeline receives a shared document — the \textbf{PRIME
DIRECTIVE} — prepended to its role-specific prompt before any LLM call.
This constitution defines four immutable laws that govern all agents:

\begin{enumerate}[noitemsep,leftmargin=*]

\item \textbf{First-Principles Derivation.}
Models must derive performance from physical and architectural constraints, never
by scaling or fitting reported numbers.
The distinction is illustrated with explicit code examples (Figure~\ref{fig:prime-directive}).
The forbidden pattern takes a paper's reported throughput and divides by the number
of units to get per-unit throughput, then multiplies back — validating the paper's
claims against scaled versions of the same claims.
The required pattern derives throughput from hardware parameters (core count,
frequency, memory bandwidth), then \emph{compares} the result to the paper's
reported values as an independent check.

\item \textbf{Calibration (The Truth Overlay).}
Agents must extract specific data points from the paper's figures, tables, and text
and plot them as scatter markers on top of the theoretical curves.
If a claimed data point sits above the theoretical curve, the paper is asserting
performance that its described mechanism cannot support — this is the ``magic gap.''

\item \textbf{Self-Containment.}
The final \texttt{model.py} must run as \texttt{python model.py} with no arguments,
no external files, and no network access.
It generates its own synthetic workload data from distributions defined in the specification.

\item \textbf{Baseline vs.\ Proposed.}
Every analysis must model both the status quo and the new contribution side-by-side.
A proposed system analyzed in isolation — without a baseline — is scientifically
meaningless.

\end{enumerate}

The constitution is the mechanism by which Rosetta enforces scientific rigor
across heterogeneous agents.
Without it, agents routinely produce plausible-looking but methodologically invalid
models — a finding confirmed by ablation.

\begin{figure}[t]
  \centering
  \fbox{\parbox{\columnwidth}{\small
    \textbf{Forbidden (circular):}
    \texttt{per\_vcu = paper\_reported\_20vcu / 20; throughput\_8vcu = per\_vcu * 8}\\[4pt]
    \textbf{Required (first principles):}\\
    \texttt{mpix\_per\_core = pixels * fps / 1e6} \quad \textit{(from Section 3.1)}\\
    \texttt{compute\_bound = mpix\_per\_core * cores\_per\_vcu}\\
    \texttt{theoretical\_max = min(compute\_bound, memory\_bound)}\\
    \texttt{efficiency = paper\_claimed / theoretical\_max} \quad \textit{(15\% gap!)}
  }}
  \caption{The PRIME DIRECTIVE's first-principles law illustrated.
    The forbidden pattern validates paper results against scaled versions of
    themselves — no independent derivation.
    The required pattern derives performance from hardware specs and then
    \emph{compares} the result to the paper's reported value.}
  \label{fig:prime-directive}
\end{figure}

\subsection{The Best-of-$N$ Ensemble}
\label{sec:ensemble}

LLM outputs are stochastic: the same pipeline on the same paper produces different
specifications, models, and interpretations across runs.
Some runs surface sharp insights; others miss key bottlenecks or make incorrect
assumptions.
Drawing on recent findings in LLM self-consistency and Best-of-$N$
reasoning~\cite{cobbe2021training,brown2024large},
Rosetta runs the full three-phase pipeline $N{=}3$ times on each paper and uses
a selector agent to choose the best run.

The selector executes \texttt{python model.py} for each run to confirm it runs
without errors, then evaluates three criteria in priority order:
\textbf{Correctness} (primary; a model that crashes or is trivially wrong is
disqualified regardless of other merits),
\textbf{Insight Quality} (primary differentiator; does the run identify concrete
bottlenecks, quantify magic gaps, and go beyond restating the paper?),
and \textbf{Completeness} (secondary; are all subsystems modeled, are calibration
targets included?).
The selector produces a per-run evaluation report and a final recommendation with
explicit justification.

\myparagraph{Ensemble quality in practice.}
Across our case study papers, the three ensemble runs diverged \emph{meaningfully}
— not merely as minor numerical variations, but in modeling philosophy and
interpretation quality.
Consider two failure modes we observed.
In one run on the Anton paper, the model predicted performance
5--6$\times$ higher than the paper claimed, yet the interpretation document
declared this a validation success — a failure of critical thinking
that the selector identified and rejected.
In another run on the CraterLake paper, a bug in the baseline formulation made
the proposed-vs-baseline comparison meaningless, despite the individual cost
functions being correct.

The selector's primary differentiator — insight quality — consistently surfaces
the run that treats unexplained gaps as \emph{signals to flag} rather than
problems to rationalize away.
In the Darwin analysis, only one of three runs correctly noted that the headline
``15{,}000$\times$ speedup'' applies only to ONT\_2D reads and drops to
1{,}244$\times$ for ONT\_1D reads due to error-rate cascade — a nuance buried in
the paper's details.

Notably, there is no consistent bias toward which run number wins: across our
papers, runs 1, 2, and 3 each won in roughly equal proportion.
This confirms that the ensemble is not redundant — each run is a genuinely
independent draw from the model's hypothesis space.
The ensemble also provides containment when the verify-repair loop does not fully
converge: in one case study, a flaw in the
F1+ baseline model survived three spec critique rounds and was faithfully
propagated into the model by the functional verifier — which checks code-to-spec
fidelity, not spec-to-physics correctness.
The selector identified the affected run and chose an alternate with a sound
baseline derivation.

\subsection{Agent Inventory and Implementation}
\label{sec:agents}

Table~\ref{tab:agents} summarizes the eight agents in the system.
The pipeline is implemented as a fully automated Python program that
orchestrates LLM API calls, parses structured outputs, manages file I/O,
and executes generated code --- no interactive chat interface is involved.
Six of the eight agents come in generator/verifier or generator/repair
pairs — a deliberate separation of concerns where no agent ever evaluates
its own work.
All agents run at temperature 0.5; the ensemble selects the best output across
three full pipeline runs (Section~\ref{sec:ensemble}).
Ensemble runs are executed sequentially rather than in parallel to simplify
debugging and resource management; total runtime remains practical.

\myparagraph{Model choice.}
All experiments were run with Claude Opus~4.5 (training data cutoff May 2025).
In our experiments, Gemini-2.5, Gemini-3, and ChatGPT series models provided
inferior results on our task.
Frontier model comparison is orthogonal to our contributions and is an area
for future work; we expect that other models will improve over time and that
Rosetta's architecture will be able to leverage those improvements as they
arise.

Output token budgets follow a consistent two-tier design: generator and repair
agents start at 16{,}384 tokens, while verifier agents start at 8{,}192 tokens,
reflecting that verification requires shorter structured responses than
generation.
All agents share a hard cap of 64{,}000 tokens.
If an agent terminates due to a context-length error, it retries up to three
times with a 1.5$\times$ budget multiplier per attempt (16{,}384 $\to$ 24{,}576
$\to$ 36{,}864 $\to$ 55{,}296, capped at 64{,}000), accommodating long papers
and complex derivations without permanently inflating the budget.

\begin{table}[t]
  \centering
  \caption{Rosetta agent inventory. All agents receive the PRIME DIRECTIVE
    prepended to their role-specific prompt.}
  \label{tab:agents}
  \small
  \begin{tabular}{lll}
    \toprule
    \textbf{Agent} & \textbf{Phase} & \textbf{Role} \\
    \midrule
    Prof.\ Aximo      & 1 & Spec generator      \\
    Dr.\ Logic        & 1 & Spec verifier       \\
    Prof.\ Aximo (CM) & 1 & Spec repair         \\
    Vector            & 2 & Model implementer   \\
    QA Engineer       & 2 & Functional verifier \\
    PI                & 2 & Directive verifier  \\
    Vector (DM)       & 2 & Model repair        \\
    Tech.\ Communicator & 3 & Interpretation    \\
    \bottomrule
  \end{tabular}
\end{table}


\subsection{Why Naive Design Fails}
\label{sec:why_naive_fails}

Each design decision in the pipeline addresses a specific, empirically
observed failure mode.
The examples below are from Case Study~1.
To prevent evaluation bias, the papers used during Rosetta's initial
development (Cambricon-SR~\cite{cambricon-sr} and F1~\cite{F1-paper})
were strictly excluded from our case studies; instead, they served as
the formative testbeds that revealed the failure modes described below
and helped develop and refine Rosetta's current design.

\myparagraph{Constitution and spec verification.}
Without the PRIME DIRECTIVE, LLMs default to circular reasoning: an
unguided CraterLake run back-calculated efficiency from the paper's
claimed 11.2$\times$ speedup and predicted an impossible
465$\times$.
Careful prompting alone is insufficient; the constitution provides
concrete code-level anti-patterns that all agents enforce.
Even with the constitution, spec-level errors can survive generation and
propagate faithfully into code.
A flawed CraterLake spec assigned CraterLake-specific hardware (the CRB
unit) to the F1+ baseline, which does not have one;
the resulting \texttt{model.py} passed code verification because it
perfectly matched its buggy spec.
A dedicated spec verification loop that reads against the original paper
--- before any code is written --- is the only mechanism that catches
this class of error.

\myparagraph{Dual verification and ensemble.}
Functional correctness and scientific validity are orthogonal.
In one Anton run, the code passed
functional verification because it matched its spec, but the PI agent
caught that the primary computation used a table of paper results --- a
first-principles violation baked into the spec itself.
A single verifier checking both axes reliably deprioritizes one.
Finally, LLM stochasticity makes single-run output unreliable: on NeuRex,
Run~1 predicted a $3{\times}$
\emph{slowdown}; Run~2 crashed yet hallucinated an interpretation; only
Run~3 produced a sound derivation.
On Darwin, only one of three runs correctly decomposed the
15{,}000$\times$ headline speedup into per-workload components.
The best-of-$N$ ensemble with a selector agent exploits this variance
to reliably extract the most analytically sound output.
Advanced single-agent environments (e.g., Claude Code) offer execution
and error-repair loops but collapse all of these safeguards: a single
agent critiquing its own spec suffers from confirmation bias, cannot
perform orthogonal dual verification, and precludes the stochastic
variance the ensemble exploits.
Multi-agent separation is not scaffolding; it is the core mechanism
guarding against self-consistent but flawed reasoning.


\section{Evaluation Methodology}
\label{sec:eval-methodology}

Rosetta outputs mechanistic understanding, not numerical predictions alone.
Evaluating it faces the same challenge as evaluating simulation
frameworks~\cite{li2009mcpat,binkert2011gem5,accelsim} and benchmark
suites~\cite{clearingclouds}: no single automated metric captures
whether an analytical artifact is correct, insightful, and useful.
We evaluate across three complementary tracks --- correctness,
generalization, and practitioner adoption as detailed below across three
types of case studies (CS).

CS1 (Section~\ref{sec:cs1}) evaluates \emph{correctness and insight}:
\papercount{} landmark papers spanning diverse domains, each assessed by
an independent human evaluator using a structured rubric --- analogous to
validating a simulator against known microarchitectural behavior.
CS2 (Section~\ref{sec:cs2}) tests \emph{generalization}: Rosetta runs
on \cstwopaper{} uncurated ISCA~2025 and HPCA~2026 papers, scored by
the automated selector, to confirm that CS1's curated selection does not
flatter the system.
CS3 (Section~\ref{sec:cs3}) measures \emph{practitioner value}: we give
Rosetta to \csthreecount{} active research groups whose authors have
simulation ground truth and design intent absent from the PDF, and
collect self-evaluations including concrete impact on the paper under
development.

All results reported here are from Rosetta running in
\emph{fully autonomous mode} (a single launch of the multi-agent
pipeline with only the paper PDF as input, zero human intervention,
no post-hoc editing) to establish a \emph{lower bound on capability.}
In practice, the generated artifacts are self-contained and well-commented;
researchers can load them into any AI-assisted coding environment
(e.g., Claude Code, Cursor) alongside the PDF and iterate directly.


\myparagraph{Frontier model training data.}
The CS1 landmark papers (published 2007--2023) are likely in Claude
Opus~4.5's training data (cutoff May 2025).
However, Rosetta's task --- deriving a structured mathematical
specification and executable model cognizant of hardware constraints is
distinct from recall or summarization. CS2's ISCA~2025 and HPCA~2026 papers were published at or after the
training cutoff, and CS3's papers are unpublished work no frontier model
has seen. CS1 is thus the most conservative evaluation condition; CS2 and CS3 provide progressively stronger completely contamination-free evidence. The PRIME DIRECTIVE explicitly prohibits using reported results as formula inputs, so even memorized paper content cannot enter the derivation.

In the main text we present synthesized findings and representative deep dives. The full generated corpus will be released as a companion artifact.


\section{Case Study 1: Correctness and Insight}
\label{sec:cs1}
As described in Section~\ref{sec:eval-methodology}, CS1 evaluates correctness and insight depth: \emph{does Rosetta produce analytical models that help an outsider understand a paper's performance claims --- correctly identifying bottlenecks, surfacing implicit assumptions, and quantifying gaps between the described mechanism and the reported results?} We curated a set of architecture papers spanning diverse problem domains, asked experienced architecture researchers to evaluate Rosetta's outputs against both the paper and their own understanding using a structured rubric, and analyzed the results.

\subsection{Experimental Setup}
\label{sec:cs1-setup}

\myparagraph{Paper selection.}
We selected \papercount{} papers from ISCA, ASPLOS, and MICRO spanning
seven domains (Table~\ref{tab:cs1-scores-compact}), chosen to cover a
wide difficulty spectrum: from papers with explicit analytical formulas
(e.g., CraterLake) to those supported primarily by simulation
(e.g., Warehouse-scale Video), with citation counts ranging from 39 to
over 1{,}000.

\myparagraph{Evaluation protocol.}
The automated selector chose the best of three ensemble runs per paper
(Section~\ref{sec:ensemble}).
Each human evaluator\footnote{We use ``human evaluator'' throughout to
distinguish the independent researchers who assessed Rosetta's outputs
from the automated selector and verifier agents within the pipeline.}
then read or re-read the paper, examined all three output artifacts, and
completed the rubric provided (3--7 hours per paper).
Evaluators were architecture researchers with broad hardware-systems
competence (self-rated 1--3/5 on domain-specific expertise), recruited
independently of Rosetta, given no briefing on what constitutes
``good'' output, and instructed to be as critical as appropriate.
They completed rubrics independently with no access to paper code,
or simulators, and are not co-authors on this paper.

\myparagraph{Rubric structure.}
This case study's rubric covers four dimensions:

\begin{itemize}[noitemsep,leftmargin=*]
\item \textbf{Specification quality} (Part 1): correctness of core relationship
  extraction, richness and completeness (1--5), hallucination check, and
  quality of the plain-English interpretation (1--5).
\item \textbf{Model quality} (Part 2): execution, spec fidelity, baseline vs.\
  proposed comparison, calibration overlay, self-containment, code clarity (1--5).
\item \textbf{Insight value} (Part 3): parametric understanding (1--5), assumption
  surfacing (1--5), insight beyond reading the paper alone (1--5), and gap
  identification.
\item \textbf{Overall assessment} (Part 4): overall usefulness (1--5), best
  concrete output, biggest failure, and whether the Evaluator would use Rosetta
  again on papers in this area.
\end{itemize}

Parts~1d, 3a--3c, and 4a require the most evaluator judgment and show the greatest variance across papers.



\subsection{NeuRex (Neural Rendering) Deepdive}
\label{sec:cs1-neurex-main}
We begin with a look into one of the evaluator reports to illustrate the depth and nature of the insights produced by Rosetta, and then synthesize across all \papercount{} papers to identify recurring patterns and conditions that predict when Rosetta adds the most value.

NeuRex~\cite{TODO-neurex} is a custom hardware accelerator for neural radiance
field (NeRF) rendering (ISCA 2023) that achieves $9.17{\times}$ speedup over
an edge GPU and $2.88{\times}$ over a server GPU using restricted hashing to
shrink the multi-resolution hash table working set from 2~MB to 32~KB.
It earned the highest scores across all CS1 dimensions (Spec, Model, Insight,
Overall all 5/5).
To illustrate exactly what Rosetta generates, we show concrete excerpts from
the NeuRex artifacts before discussing the findings.

\begin{figure}[t]
  \centering
  \includegraphics[width=\columnwidth]{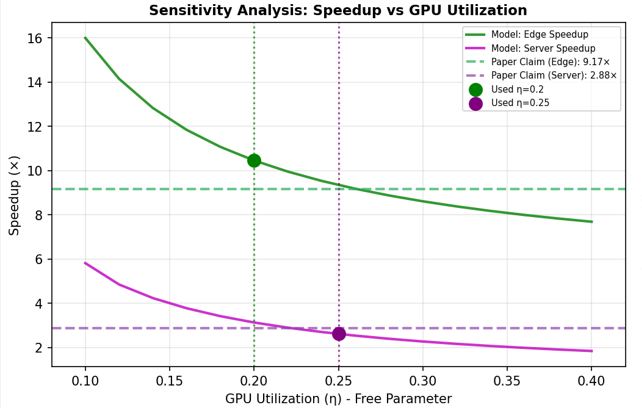}
  \caption{Sensitivity analysis from the NeuRex model: predicted speedup
    vs.\ GPU utilization $\eta$ for the Edge (Xavier NX, green) and Server
    (RTX~3070, purple) configurations.
    Dashed lines show the paper's claimed speedups.}
  \label{fig:neurex-sensitivity}
\end{figure}

\myparagraph{What Rosetta generates: artifact excerpts.}
Figure~\ref{fig:neurex-spec-excerpt} shows a representative excerpt from
the generated specification: the hardware parameter table for the
NeuRex accelerator and baseline GPUs, with every value traced to a specific
paper section.
Figure~\ref{fig:neurex-code-excerpt} shows the corresponding
\texttt{model.py} implementation of the GPU bandwidth-efficiency derivation ---
the core first-principles calculation that drives the model's predictions.
The full specification is 643 lines; the full model is 1{,}177 lines of
self-contained Python. Figure~\ref{fig:neurex-sensitivity} is 
one of the graphs produced by the model (details shortly).
All three artifacts (spec, model, interpretation) are generated autonomously typically in 30 to 40 minutes.

\begin{figure}[t]
\fbox{\parbox{0.97\columnwidth}{\footnotesize
\textbf{From \texttt{SPECIFICATION.md} --- Hardware Parameters (excerpt):}\\[4pt]
\begin{tabular}{@{}llll@{}}
\textbf{Variable} & \textbf{Edge} & \textbf{Server} & \textbf{Source} \\
$N_{IGU}$ (Index Gen.\ Units) & 8 & 64 & \S 4.4, Tab 4 \\
$S_{GC}$ (Grid cache) & 64 KB & 64 KB & Tab 4, \S 6.4 \\
$S_{SB}$ (Subgrid buffer) & 128 KB & 128 KB & Tab 4 \\
$N_{SA}$ (Systolic arrays) & 1 & 16 & \S 5, Tab 4 \\
$f_{clk}$ & 1 GHz & 1 GHz & \S 5 \\
\end{tabular}\\[4pt]
\textbf{Baseline GPU parameters:}\\[2pt]
\begin{tabular}{@{}llll@{}}
& \textbf{Xavier NX} & \textbf{RTX 3070} & \textbf{Source} \\
$S_{L2}$ (L2 cache) & 256 KB & 4 MB & Sec 3.4, public spec \\
$BW_{mem}$ & 51.2 GB/s & 448 GB/s & Public spec \\
\end{tabular}\\[4pt]
\textit{BW Eff (GPU):}
$\eta_{BW,GPU} = B_{entry} / B_{cacheline} = 4/64 = 0.0625$\\
\hspace*{2em}\textit{Source: Sec 3.4, ``each hash entry access only uses 4 out of 64 bytes.''}
}}
\caption{Excerpt from the generated \texttt{SPECIFICATION.md} for NeuRex.
  Every parameter is traced to a paper section; the bandwidth efficiency
  is derived from hardware fundamentals.}
\label{fig:neurex-spec-excerpt}
\end{figure}

\begin{figure}[t]
\fbox{\parbox{0.97\columnwidth}{\footnotesize
\textbf{From \texttt{model.py} --- GPU bandwidth efficiency (excerpt):}\\[4pt]
\texttt{def compute\_gpu\_bandwidth\_efficiency(}\\
\texttt{~~~~cacheline\_bytes, bytes\_per\_entry):}\\
\texttt{~~\textquotedbl\textquotedbl\textquotedbl From Sec 3.4: `each hash entry access}\\
\texttt{~~only uses four out of 64 bytes'\textquotedbl\textquotedbl\textquotedbl}\\
\texttt{~~return bytes\_per\_entry / cacheline\_bytes}\\
\texttt{~~\# 4/64 = 0.0625}\\[4pt]
\texttt{\# Cache fit analysis (first principles):}\\
\texttt{\# Xavier NX: 2 MB table > 256 KB L2 -> misses}\\
\texttt{\# RTX 3070:~~2 MB table < 4 MB L2~~~-> fits}\\
\texttt{table\_fits = gpu.l2\_cache >= hash.table\_size}\\[4pt]
\texttt{\# Speedup = T\_gpu / T\_neurex (NOT calibrated)}\\
\texttt{\# eta\_util swept in [0.10, 0.40]}
}}
\caption{Excerpt from the generated \texttt{model.py} for NeuRex.
  The code derives performance from hardware parameters; the paper's
  reported results appear only as scatter overlays for comparison, never
  as formula inputs.}
\label{fig:neurex-code-excerpt}
\end{figure}

\myparagraph{Key analytical finding.}
Rosetta's primary contribution for NeuRex is a first-principles derivation
of why GPUs struggle with multi-resolution hash encoding:
\[
\eta_{\mathrm{BW,GPU}} = \frac{B_{\mathrm{entry}}}{B_{\mathrm{cacheline}}}
= \frac{4\;\mathrm{bytes}}{64\;\mathrm{bytes}} = 6.25\%
\]
Each hash table entry stores two 16-bit features, but GPU memory fetches
64-byte cache lines, wasting 60 of every 64 bytes.
This structural inefficiency explains why NeuRex-Edge ($9.17{\times}$) outpaces
NeuRex-Server ($2.88{\times}$) by such a wide margin: the Xavier NX's 256~KB L2
cache cannot hold the 2~MB hash table, forcing every lookup to LPDDR4 main
memory, while the RTX~3070's 4~MB L2 already partially solves the problem
NeuRex is designed to address.
The model sweeps GPU utilization $\eta \in [0.10, 0.40]$ without calibrating
to the paper's claims; both speedup figures fall within the predicted band
(Figure~\ref{fig:neurex-sensitivity}). The human evaluator captured the mode of value precisely:
\begin{quote}
\textit{``Makes plain from base principles how the new accelerator outpaces
GPUs.
It provides upfront analysis of the paper's claims and exposes the
gaps/withheld information of the paper.''}
\end{quote}

\subsection{Aggregate Results}
\label{sec:cs1-aggregate}

Table~\ref{tab:cs1-scores-compact} summarizes evaluation scores across the \papercount{} papers, which we analyze further below.

\begin{table}[t]
  \centering
  \caption{CS1 scores (\papercount{} papers, 1--5 scale).
    \textbf{S}=Spec+Interp, \textbf{M}=Model, \textbf{R}=Richness,
    \textbf{P}=Parametric, \textbf{A}=Assumption surfacing,
    \textbf{I}=Insight${>}$Paper, \textbf{O}=Overall.
    \textbf{?}=Use Again (D=Definitely, P=Probably, U=Unsure).}
  \label{tab:cs1-scores-compact}
  
  \setlength{\tabcolsep}{2.5pt}
  \begin{tabular}{@{}llccccc w{c}{1em} w{c}{1em} w{c}{1em}@{}}
    \toprule
    \textbf{Paper} & \textbf{Domain} & \textbf{S} & \textbf{M} & \textbf{R} & \textbf{P} & \textbf{A} & \textbf{~~I} & \textbf{~~O} & \textbf{~~?} \\
    \midrule
    BASALISC~\cite{TODO-basalisc}     & FHE        & 5 & 5 & 4 & 4 & 4 & \textbf{4} & \textbf{5} & \textbf{D} \\
    Darwin~\cite{TODO-darwin}         & Genomics   & 4 & 5 & 5 & 3 & 2 & \textbf{5} & \textbf{5} & \textbf{D} \\
    CraterLake~\cite{TODO-craterlake} & FHE        & 4 & 3 & 4 & 4 & 1 & \textbf{4} & \textbf{5} & \textbf{P} \\
    Warehouse~\cite{TODO-warehouse}   & Video      & 4 & 4 & 3 & 5 & 3 & \textbf{3} & \textbf{4} & \textbf{U} \\
    Anton~\cite{TODO-anton}           & Mol.Dyn.   & 4 & 5 & 4 & 2 & 3 & \textbf{4} & \textbf{4} & \textbf{P} \\
    GenAx~\cite{TODO-genax}           & Genomics   & 4 & 4 & 4 & 4 & 3 & \textbf{4} & \textbf{5} & \textbf{P} \\
    Genesis~\cite{TODO-genesis}       & Genomics   & 5 & 4 & 4 & 5 & 3 & \textbf{4} & \textbf{5} & \textbf{P} \\
    SeGraM~\cite{TODO-segram}         & Genomics   & 3 & 2 & 3 & 3 & 1 & \textbf{3} & \textbf{4} & \textbf{P} \\
    GZKP~\cite{TODO-gzkp}             & ZK Proofs  & 3 & 5 & 4 & 3 & 3 & \textbf{3} & \textbf{4} & \textbf{P} \\
    PipeZK~\cite{TODO-pipezk}         & ZK Proofs  & 4 & 3 & 3 & 4 & 4 & \textbf{4} & \textbf{5} & \textbf{P} \\
    NeuRex~\cite{TODO-neurex}         & Neur.Rend. & 5 & 5 & 4 & 4 & 4 & \textbf{5} & \textbf{5} & \textbf{D} \\
    MD-Pipe~\cite{TODO-mdpipe}        & Mol.Dyn.   & 4 & 4 & 4 & 3 & 3 & \textbf{4} & \textbf{5} & \textbf{P} \\
    \bottomrule
  \end{tabular}
\end{table}

\myparagraph{Spec and interpretation quality is the most reliable output.}
Across all \papercount{} evaluations, Spec+Interpretation quality scores range from
3 to 5, with ten of twelve papers at 4 or above.
Human evaluators found that Rosetta reliably identifies the correct hardware parameters,
constructs analytically reasonable cost functions, and produces an interpretation
document that genuinely helps a newcomer orient to the paper.
One human evaluator noted: ``I found myself using the interpretation to help guide me through
a re-read of the paper.''
Zero expert evaluations reported significant hallucinations; four reported minor ones
(CraterLake: assumed register file port counts and an arbitrary F1+ bandwidth value;
Darwin: a bandwidth unit misinterpretation in the de~novo throughput path;
Warehouse: assumed combined read/write accounting;
SeGraM: an HBM channel--accelerator topology assumption)
that human evaluators considered inconsequential to the core derivation.

\emph{\textbf{Key takeaway:} The specification and interpretation are Rosetta's most reliable outputs, scoring 4+ on 10 of 12 papers with zero significant hallucinations.}

\myparagraph{Model quality is strong but bounded by spec quality.}
Model quality scores range from 2 to 5, with five papers at 5/5
(BASALISC, Darwin, Anton, GZKP, NeureX).
In one confirmed case, the model corrects a minor spec-level error:
Darwin's \texttt{model.py} fixes an incorrect de~novo throughput formula that
propagated from the specification.
In the opposite direction, Anton's incorrect $T_{\text{long-range}}$ formula is
faithfully reproduced from the spec into the model, illustrating that the pipeline's
verify step catches scientific invalidity but does not independently audit
arithmetic against the paper.
The 3/5 scores for CraterLake and PipeZK both reflect spec-level gaps propagated
into code: for CraterLake, the F1+ baseline incorrectly used CraterLake's CRB unit
cycles when F1+ has no such hardware;
the single 2/5 (SeGraM) reflects an inadequate baseline model: the BitAlign
accelerator derivation correctly follows Algorithm~1 line-by-line, and the
100$\times$ end-to-end latency gap is correctly identified as the central
analytical finding, but the CPU comparison system is modeled as a generic
processor rather than one implementing the MinSeed and BitAlign algorithms
that SeGraM accelerates — a gap the paper's own sparse baseline description
makes difficult to close from the PDF alone; and flagged in the SPEC critique,
and observed in INTERPRETATION.
This confirms a key design constraint: model quality is bounded below by spec
quality — a sound model cannot be built on an insufficient specification.

\emph{\textbf{Key takeaway:} Model quality (2--5) is never higher than spec quality; the verify-repair loop's primary leverage point is the specification phase.}

\myparagraph{Insight value is the most discriminating dimension.}
The ``insight beyond paper'' score varies most widely (3--5) and is the strongest
predictor of overall usefulness.
Rosetta adds most value when: (i)~the paper presents sparse mathematical
detail but rich hardware description, forcing assumptions into the open;
(ii)~the paper reports performance at a single operating point, and the parametric
model reveals behavior across a design space.
The CraterLake $L$-dependence sweep is the clearest example of (ii): the paper
reports speedup at fixed benchmarks, but Rosetta produced a continuous speedup
vs.\ $L$ curve showing 4$\times$ at low $L$ (shallow FHE) to 51$\times$ at high
$L$ (deep FHE) — a result not available from the paper's tables alone.
NeureX and Darwin are the two papers reaching 5/5 insight; in both cases the paper
describes dense hardware mechanisms whose parametric interaction the model resolves
into a navigable design-space map.

\emph{\textbf{Key takeaway:} Insight value is highest when papers have rich hardware descriptions but sparse explicit math, and when parametric sweeps reveal design-space behavior hidden in point results.}

\myparagraph{Where insight is bounded.}
Three papers score 3/5 on insight, each for a distinct structural reason.
Warehouse-scale Video illustrates a \emph{parameter-withholding ceiling}: key
configuration values are withheld for proprietary reasons, preventing Rosetta
from explaining the 6.5$\times$ gap between theoretical peak and observed
throughput.
The Evaluator credited the model for correctly reaching this conclusion and flagging
it: ``the model knows why the model is lean --- I don't think a human could do any
better.''
GZKP illustrates a \emph{component-divergence ceiling}: NTT results are confirmed
analytically, but the MSM component model diverges substantially from empirical
results, ``calling more questions about itself than the paper it is meant to
reflect.''
SeGraM illustrates a \emph{missing-baseline ceiling}: without a model for the
CPU baseline that implements MinSeed and BitAlign, the comparative insight is
limited to confirming the accelerator's own scalability properties rather than
explaining the relative speedup.
Across all three, Rosetta correctly diagnoses the ceiling rather than
producing spuriously confident output.

\emph{\textbf{Key takeaway:} When insight is bounded, Rosetta identifies why --- parameter withholding, component divergence, or missing baselines --- rather than spuriously confident output.}

\myparagraph{When Does Rosetta Add Most Value?}
\label{sec:cs1-synthesis}
Across all \papercount{} evaluations, Rosetta's insight value is a function
of the \emph{paper's mathematical disclosure posture}, not of Rosetta's
intrinsic capability.
Crucially, the three papers scoring 3/5 on insight --- Warehouse-scale Video,
SeGraM, and GZKP --- share a common characteristic: Rosetta correctly
diagnosed its own ceiling rather than producing spuriously confident output,
a property that makes its confirmed findings credible.
Table~\ref{tab:cs1-findings} catalogs the eight recurring finding types
observed across the \papercount{} papers; four conditions consistently predict
high insight value: \textbf{(1)}~rich hardware description with sparse explicit math
(forces assumptions into the open); \textbf{(2)}~single-point results whose
latent parametric structure the model reveals; \textbf{(3)}~a gap between
first-principles prediction and claimed results that characterizes rather than
condemns; and \textbf{(4)}~formalization as standalone value, where the specification
delivers insight even when the quantitative model is imperfect.

\begin{table*}[t]
  \centering
  \caption{Recurring finding types across CS1, with representative papers and concrete outputs.}
  \label{tab:cs1-findings}
  \small
  \setlength{\tabcolsep}{4pt}
  \begin{tabular}{p{4.3cm} p{2.2cm} p{9cm}}
    \toprule
    \textbf{Finding type} & \textbf{Papers} & \textbf{What Rosetta produced} \\
    \midrule
    Structural hardware bottleneck from first principles
      & NeuRex, Genesis
      & 6.25\% BW efficiency ($4/64$ bytes/cache-line);
        2.08$\times$ achieved is within 1.4\% of Amdahl ceiling \\
    \hline
    Parametric sweep reveals hidden design-space behavior
      & CraterLake, NeuRex, PipeZK
      & $L$-dependence curve ($4{\times} \to 51{\times}$);
        $\eta$-sweep bracketing claimed speedup without calibration;
        degradation from $197{\times}$ to $30{\times}$ across problem sizes \\
    \hline
    Module-to-system gap closed analytically
      & PipeZK, Genesis
      & Amdahl prediction $5.2{\times}$ vs.\ paper's $5.8{\times}$;
        ceiling identifies true performance floor \\
    \hline
    Formalization as standalone value (model imperfect)
      & PipeZK
      & 5/5 Overall despite 3/5 Model;
        qualitative paper statements converted to quantified boundaries \\
    \hline
    Component-level audit: confirmation and divergence
      & GZKP
      & NTT $O(N\log N)$ scaling verified;
        MSM sign-flip inverts speedup prediction \\
    \hline
    Model corrects a spec-level paper error
      & Darwin
      & Incorrect de~novo throughput formula caught in implementation phase \\
    \hline
    Paper withholds quantitative data; model makes claims checkable
      & NeuRex, Warehouse
      & $\eta$-sweep without calibration; proprietary ceiling correctly diagnosed \\
    \hline
    Missing-baseline ceiling correctly diagnosed
      & SeGraM
      & 100$\times$ latency gap attributed to undocumented optimizations;
        generic-CPU baseline limits comparative insight \\
    \bottomrule
  \end{tabular}
\end{table*}

\subsection{CAAM-Bench: CompArch Analytical Model Benchmark}
\label{sec:caam-bench}

No standard benchmark exists for evaluating tools that automatically generate
analytical models from architecture papers, making reproducible comparison
across approaches or over time impossible.
We release the CS1 evaluation set as \textbf{CAAM-Bench} (CompArch Analytical
Model Benchmark) to fill this gap.
CAAM-Bench has three components:

\begin{itemize}[noitemsep,leftmargin=*]
\item \textbf{12 landmark architecture papers} spanning seven domains
  (Table~\ref{tab:cs1-scores-compact}), selected to cover a wide difficulty
  spectrum from papers with explicit analytical formulas to simulation-heavy
  works, with citation counts from 39 to over 1{,}000.

\item \textbf{Reference artifacts}: for each paper, a human-vetted
  \texttt{SPECIFICATION.md} and \texttt{model.py} produced by Rosetta
  and confirmed by an independent domain-expert evaluator.
  These serve as a quality baseline --- a concrete target for future tools
  to match or exceed.

\item \textbf{A scoring protocol}: the four-dimension CS1 rubric
  (specification quality, model quality, insight value, overall usefulness)
  applied to the spec and model produced by any tool or LLM on the same
  12 papers.
  The rubric can be applied by human evaluators or, following CS2's approach,
  by an automated scorer calibrated against the human-vetted reference.
  Rosetta's scores (Table~\ref{tab:cs1-scores-compact}) serve as the
  published baseline.
\end{itemize}

CAAM-Bench enables reproducible evaluation of future analytical model
generators by providing a fixed paper corpus, reference artifacts, and
a common scoring protocol.
The 12 papers span sufficient domain diversity --- FHE, genomics, molecular
dynamics, ZK proofs, neural rendering, and video --- that performance on
any single domain is unlikely to generalize without genuine analytical depth.
All benchmark materials are released with this paper.


\section{Case Study 2: Breadth Study}
\label{sec:cs2}

To confirm that CS1's curated selection does not flatter the system, we
ran Rosetta on 97 unfiltered papers from ISCA~2025 (80) and HPCA~2026
(17).
An independent fit judge excluded 17~papers unsuitable for first-principles
analysis (scheduling heuristics, purely algorithmic work); the remaining
\textbf{78~papers} form the reported corpus.
Across these, mean Insight Quality is 8.2/10 and mean Correctness is
7.8/10; 56\% reach Tier~A (9--10) and 91\% score 7+.
We classified outcomes into three categories:
\textbf{GAP-ONLY} (60, 77\%) where the model quantifies a discrepancy
without attributing a cause it cannot derive;
\textbf{SHOWS-NEW-INSIGHT} (12, 15\%) where the derivation surfaces a
tension with a specific claim not visible from reading alone; and
\textbf{VALIDATES} (5, 6\%) where the derivation independently confirms
the paper's result.
The median pipeline runtime is 73~minutes at $\sim$\$39/paper
(\$3{,}800 total).

\emph{\textbf{Key takeaway:} Rosetta generalizes at conference scale,
with 15\% of papers surfacing analytical tensions not visible from reading
the paper alone.}


\section{Case Study 3: Researcher Use}
\label{sec:cs3}

As described in Section~\ref{sec:eval-methodology}, CS3 tests practitioner
value: \emph{does Rosetta add value during active research?}
We gave Rosetta to six research groups with papers at premier
architecture venues (submitted, under revision, or accepted) and
collected author self-evaluations using a structured rubric
(Figure~\ref{fig:cs3-rubric}).
Authors evaluated Rosetta's output against their own simulation
ground truth and design intent --- context absent from the PDF and
unavailable to CS1's external evaluators.
Titles and venues are redacted as the papers are active
submissions\footnote{All papers target top-3 architecture or systems
venues.}.

\subsection{Researchers Want to Use Rosetta Again}
\label{sec:cs3-headline}

We do not bury the lede.
Five of six CS3 evaluators --- including those who gave the lowest overall
scores --- said ``Definitely'' when asked whether they would run Rosetta
on their next paper; the sixth said ``Unsure'' only because their paper was
a simulation study at the hardest analytical ceiling.
The impact these researchers describe goes beyond anything a single
accuracy number can capture.
SYMI's evaluator: \textit{``No reviewer would have the time to write a
simulation to validate a paper's empirical results, but with Rosetta
it's a 5-minute task.''}
SgBend's evaluator: \textit{``If we had used this from the beginning, we
might have been able to have a more profound influence on the final work
--- it also may have helped us earlier for idea development and eliminating
less promising ideas.''}
D-Com's evaluator: \textit{``Made assumptions explicit, and the generated
spec is useful.''}
The Mamba evaluator gave the only 5/5 spec richness in the CS3 set:
\textit{``Did a great job comprehending the Einsums, the fusion taxonomy,
and the 2D/1D reconfigurable array.\ Surprisingly well.''}
Even the GT evaluator, who rated overall usefulness at 2/5,
confirmed that Rosetta's bandwidth under-utilization derivation
\textit{``is very similar to our simulation results.''}

\emph{\textbf{Key takeaway:} Across six independent evaluations spanning
five different system domains, the consistent signal is adoption intent ---
not because the models are perfect, but because the formalization process
itself changes how researchers engage with their own work.}

\subsection{Other Per-Paper Technical Findings}
\label{sec:cs3-findings}

We summarize the key Rosetta-driven finding from each paper below.


\textbf{SYMI (MoE training):}
Rosetta reproduced the paper's communication-cost formulas and
independently constructed a Zipfian-distribution training simulation ---
an approach the author had started and abandoned for the same reason
(too many assumptions about the skewness parameter).
The convergence ratio matched precisely (29\% faster).
The author noted that an earlier interactive run \textit{``could have
been incorporated directly''} into the
paper.

\textbf{D-Com (LLM inference):}
SPECIFICATION.md revealed that the paper's exposition insufficiently
separated the motivational GPU analysis from the proposed custom
accelerator; the paper was updated to sharpen this
boundary.

\textbf{Mamba (SSM accelerator):}
Earned 5/5 spec richness and surfaced a terminological tension ---
whether ``fully fused'' is accurate when DRAM traffic exists at
skip-connection boundaries.

\textbf{TOPO (sparsity/RT core):}
Hardware pipeline captured correctly; workload abstractions too generic
to match simulation, establishing the dataset-dependence boundary
condition.

\textbf{SgBend (SpGEMM):}
Strongest concrete impact: Rosetta identified an IPM lookup latency
the authors' own simulator had underestimated, directly motivating a
new experiment added to the paper.


\subsection{A Graph Traversal Accelerator(GT)}
\label{sec:cs3-simcase-short}
We now highlight the graph traversal simulation study (the hardest analytical
case) 
\emph{This paper studies a graph traversal accelerator. The lead author
compared Rosetta's output against cycle-accurate simulation ground truth
not present in the PDF.
The author rated the output 2/5; we report that as ground truth.}

\myparagraph{Reading synthesis value.}
Before the performance model runs, Rosetta's reading assistant independently
flagged four analytical observations: the headline speedup over a prior FPGA
baseline includes a clock-frequency component alongside the architectural
improvement; benchmark selection was bounded by a wall-clock simulation budget,
omitting larger instances; most selected instances fit on-chip, concentrating
the off-chip scalability evidence in a single reduced-cache experiment; and
the heuristic-solver comparison involves asymmetric capability sets.
The author confirmed post-hoc that the paper was subsequently rewritten to
address exactly these concerns --- the reading synthesis had identified the
same objections that program-committee reviewers later raised in writing.

\myparagraph{Quantitative model: one confirmed finding, one error.}
The model derives baseline cycles per traversal step as $t_{step}^{base}
\approx 167$~cycles, implying memory bandwidth utilization of only $\approx
3.8\%$ --- the system is severely latency-bound, not bandwidth-bound.
The author explicitly confirmed this against simulation.
A second finding --- that the accelerator has worse vertex-data cache hit rates
than the baseline (60\% vs.\ 70\%) but better adjacency-list hit rates (85\%
vs.\ 65\%) --- was also verified against simulation and found to be wrong.
We report this as a genuine model error.
First-principles derivation closes to $9.19\times$ speedup against the paper's
$22.77\times$, a $2.48\times$ unexplained gap.

\myparagraph{Author verdict and scope implication.}
The author rated overall usefulness at 2/5 and ``Unsure'' for future use,
and offered the most precise diagnosis of the gap: not missing hardware
parameters, but the LLM's lack of algorithm-specific understanding of how
the core propagation algorithm interacts with memory access patterns.
This is a hard scope boundary: even a complete parameter checklist cannot
bridge it.
The author's most actionable suggestion: \textit{``Adding a Q\&A part with a
planning phase could help make sure the model stays on track''} --- a
pre-flight audit that enumerates which analytical inputs are and are not
recoverable from the paper alone.

\subsection{Aggregate Results}
\label{sec:cs3-synthesis}

\begin{figure}[t]
  \centering
  \small
  \setlength{\tabcolsep}{4pt}
  \begin{tabular}{p{2.5cm} p{5.5cm}}
    \toprule
    \textbf{Dimension} & \textbf{What it asks} \\
    \midrule
    \textbf{Core capture} (2a)
      & Did Rosetta identify the key analytical relationships the author considers central? \\
    \textbf{Spec richness} (2b)
      & Does the spec capture baseline + proposed, parameters, operating conditions? \\
    \textbf{Interp quality} (2e)
      & Does \texttt{INTERPRETATION.md} accurately explain the work to an outsider? \\
    \textbf{Assumption surfacing} (3a)
      & Did formalizing expose implicit assumptions the author had not previously stated? \\
    \textbf{Parametric clarity} (3b)
      & Did the model illuminate parameter sensitivity the author hadn't fully explored? \\
    \textbf{Impact on paper} (3c)
      & Did Rosetta's output change any claim, section, or experiment? \\
    \textbf{Overall usefulness} (5a)
      & Holistic author rating \\
    \textbf{Use again?} (4c)
      & Would you run Rosetta during development of your next paper? \\
    \bottomrule
  \end{tabular}

  \medskip

  \begin{tabular}{llccccc}
    \toprule
    \textbf{Paper} & \textbf{Domain} & \textbf{(2a)} & \textbf{(2b)}
      & \textbf{(2e)} & \textbf{(5a)} & \textbf{(4c)} \\
    \midrule
    $GT^{\star}$       & GraphTrav.  & Partly & 4 & 3  & 2 & Unsure    \\
    $SYMI^{\dagger}$   & MoE Sys.    & Mostly & 4 & -- & 3 & Definitely \\
    $D$-$Com^{\ddagger}$ & LLM Inf.  & Yes    & 4 & 3  & 4 & Definitely \\
    $Mamba^{\star}$    & SSM         & Yes    & 5 & 3  & 3 & Definitely \\
    $TOPO^{\ddagger}$  & Sparsity    & Partly & 3 & 3  & 3 & Definitely \\
    $SgBend^{\dagger}$ & SpGEMM      & Partly & 4 & 5  & 4 & Definitely \\
    \bottomrule
  \end{tabular}

  \caption{CS3 rubric dimensions (top) and scores (bottom).
    Scored dimensions use 1--5; categorical: Yes/Mostly/Partly/No (2a)
    and Definitely/Probably/Unsure/No (4c).
    $\star$\,Rejected, being revised;
    $\dagger$\,Accepted;
    $\ddagger$\,New submission.}
  \label{fig:cs3-rubric}
\end{figure}

Spec richness averages 4.0/5 with five of six at 4+, consistent with
CS1's finding that the specification is the pipeline's most reliable
phase.
Every CS3 evaluator named \texttt{SPECIFICATION.md} as the most valuable
output, independent of model accuracy.
Four of six evaluations produced a documented change to the paper.

\emph{\textbf{Key takeaway:} The formalization pass adds value regardless
of whether \texttt{model.py} closes numerically; four of six papers were
changed as a direct result.}

\myparagraph{The \texttt{SPECIFICATION.md} artifact is the most consistently valued output.}
Every CS3 evaluator named it as the most valuable
output, independent of model accuracy.
SYMI's spec enabled an independent rediscovery of the author's own abandoned
simulation; D-Com's surfaced a structural exposition problem in the paper;
SgBend's introduced a new analytical technique for capturing data-dependent
hardware behavior that the evaluator intends to adopt.
The formalization pass adds value regardless of whether \texttt{model.py}
closes numerically.

\emph{\textbf{Key takeaway:} Every CS3 evaluator named \texttt{SPECIFICATION.md} as the most valuable output, independent of model accuracy.}

\section{Related Work}
\label{sec:related_work}

Rosetta sits at the intersection of analytical modeling, LLM-based scientific reasoning, and multi-agent system design. We survey each area, emphasizing how prior work addresses subsets of the problem Rosetta solves end-to-end.

\myparagraph{Analytical Modeling and Simulation}
Analytical models like Roofline~\cite{williams2009roofline, ding2019instruction, ofenbeck2014applying} and Amdahl's Law~\cite{amdahl1967validity, gustafson1988reevaluating, hennessy2019computer, jain1991art,hill2008amdahl} provide fundamental performance bounds. Deep learning models~\cite{ivanov2021data, dao2022flashattention, kaplan2020scaling, narayanan2021efficient, chowdhery2023palm} apply IO-complexity and empirical scaling laws. More detailed first-principles processor~\cite{karkhanis2004firstorder, eyerman2009mechanistic} and GPU models~\cite{hong2009analytical, sim2012gpuperf, baghsorkhi2010adaptive, zhang2011quantitative, huang2014gpumech, wang2020mdm, nugteren2014gpu, kiani2018gpu, lee2022gcom} rely on manual interval and reuse-distance analysis. Others tailor models to specific workloads or hybridize with simulation~\cite{heo2020realtime, gong2020paqsim, villa2021needforspeed, arafa2021pptgpu}. However, all are manually constructed by experts. In contrast, cycle-level simulators~\cite{binkert2011gem5, luo2023ramulator2, rodrigues2011sst, sanchez2013zsim, carlson2011sniper,simplescalar} require precise hardware specifications and binaries. Rosetta occupies a unique space: it \emph{automates} the creation of analytical models directly from unstructured paper prose.

\myparagraph{LLM-Based Science and Peer Review}
LLMs increasingly assist in scientific document understanding, including QA, claim verification, and literature synthesis~\cite{taylor2022galactica, lo2020s2orc, wadden2020fact, wright2022citeworth, wang2024autosurvey, baek2024researchagent}. Beyond comprehension, systems like The AI Scientist~\cite{lu2024aiscientist}, FunSearch~\cite{romera2024mathematical}, AlphaGeometry~\cite{trinh2024solving}, and formal provers~\cite{yang2024leandojo, polu2022formal} attempt automated mathematical discovery and experimentation. LLMs are also explored for automated peer review~\cite{liang2024reviewergpt, liang2024mapping}. Rosetta fundamentally differs: it does not generate novel hypotheses, train models, or recommend paper acceptance. Instead, it operates as an analytical auditor, extracting parameters from existing text to mathematically verify claimed mechanisms against first principles.

\myparagraph{Multi-Agent Architectures and Verification}
Rosetta leverages recent advances in LLM reasoning (Chain/Tree of Thought~\cite{wei2022chain, yao2024tree} to mitigate hallucination~\cite{ji2023survey}), code generation~\cite{chen2021evaluating, roziere2024code, yao2023react, wang2024executable}, and multi-agent roles~\cite{wu2023autogen, hong2024metagpt, li2023camel}. To ensure robustness, we build upon iterative self-refinement~\cite{madaan2023selfrefine, shinn2023reflexion, huang2024large, olausson2024selfrepair, le2024codechain}, multi-agent debate/evaluation~\cite{du2024improving, chan2024chateval, zheng2023judging}, and Best-of-$N$ process verification~\cite{cobbe2021training, lightman2024lets, brown2024large}. However, while Constitutional AI~\cite{bai2022constitutional} enforces behavioral norms via RLHF, Rosetta pioneers a \emph{scientific constitution} (the PRIME DIRECTIVE) enforced via strict generator-verifier separation. Crucially, we introduce an orthogonal dual-verification step (functional vs.\ scientific validity) to prevent the specific failure mode where an LLM faithfully implements a circular, non-first-principles mathematical specification. While recent systems like PaperBanana~\cite{zhu2026paperbananaautomatingacademicillustration} orchestrate specialized agents to automate the generation and critique of academic visual illustrations, Rosetta applies a similar multi-agent decomposition to the extraction of mathematical artifacts in architecture and systems.

\section{Conclusion}
\label{sec:conclusion}

Analytical performance models explain \emph{why} performance bounds exist,
yet they rarely accompany architecture papers.
Rosetta automates this: given a PDF, it produces a mathematical
specification, executable Python model, and plain-English interpretation
with zero human intervention, using a scientific constitution, dual
verification, and a best-of-$N$ ensemble to address the failure modes of
naive LLM-based generation. Across three tracks, Rosetta demonstrates correctness (10 of 12 CS1
papers at 4--5/5 specification quality, zero significant hallucinations),
generalization (56\% Tier~A across 78 unfiltered conference papers), and
practitioner value (five of six CS3 researchers said ``Definitely'' for
future use).
The consistent finding is that formalization itself is the primary value:
it surfaces implicit assumptions, flags missing parameters, and changes how
researchers engage with their own work.

We release Rosetta open-source together with \textbf{CAAM-Bench}, the
first benchmark for evaluating analytical model generators on architecture
papers, comprising 12 human-vetted paper--artifact pairs and a
four-dimension rubric to enable reproducible comparison of future tools. More broadly, Rosetta points toward a new community norm: architecture papers accompanied by a generated spec and model, making every performance claim independently auditable by any reader, without access to the authors' simulation infrastructure.


\bibliographystyle{IEEEtranS}
\bibliography{main,references,references_analytical_models,refs_cs1}

@misc{sankaralingam2026computerarchitecturesalphazeromoment,
      title={Computer Architecture's AlphaZero Moment: Automated Discovery in an Encircled World},
      author={Karthikeyan Sankaralingam},
      year={2026},
      eprint={2604.03312},
      archivePrefix={arXiv},
      primaryClass={cs.AR},
      url={https://arxiv.org/abs/2604.03312},
}

@misc{gupta2026archagentagenticaidrivencomputer,
      title={ArchAgent: Agentic AI-driven Computer Architecture Discovery},
      author={Raghav Gupta and Akanksha Jain and Abraham Gonzalez and Alexander Novikov and Po-Sen Huang and Matej Balog and Marvin Eisenberger and Sergey Shirobokov and Ngân Vũ and Martin Dixon and Borivoje Nikolić and Parthasarathy Ranganathan and Sagar Karandikar},
      year={2026},
      eprint={2602.22425},
      archivePrefix={arXiv},
      primaryClass={cs.AI},
      url={https://arxiv.org/abs/2602.22425},
}

@String{Computing = "Computing" }

@String{Computer = "{IEEE} Computer" }

@String{Academic = "Academic Press" }

@String{Springer = "Springer-Verlag" }

@ArtifactSoftware{R,
    title = {R: A Language and Environment for Statistical Computing},
    author = {{R Core Team}},
    organization = {R Foundation for Statistical Computing},
    address = {Vienna, Austria},
    year = {2019},
    url = {https://www.R-project.org/},
}

@ARTICLE{simplescalar,
  author={Austin, T. and Larson, E. and Ernst, D.},
  journal={Computer}, 
  title={SimpleScalar: an infrastructure for computer system modeling}, 
  year={2002},
  volume={35},
  number={2},
  pages={59-67},
  doi={10.1109/2.982917}}

@inproceedings{accelsim,
author = {Khairy, Mahmoud and Shen, Zhesheng and Aamodt, Tor M. and Rogers, Timothy G.},
title = {Accel-sim: an extensible simulation framework for validated GPU modeling},
year = {2020},
isbn = {9781728146614},
publisher = {IEEE Press},
url = {https://doi.org/10.1109/ISCA45697.2020.00047},
doi = {10.1109/ISCA45697.2020.00047},
booktitle = {Proceedings of the ACM/IEEE 47th Annual International Symposium on Computer Architecture},
pages = {473–486},
numpages = {14},
location = {Virtual Event},
series = {ISCA '20}
}

@misc{zhu2026paperbananaautomatingacademicillustration,
      title={PaperBanana: Automating Academic Illustration for AI Scientists}, 
      author={Dawei Zhu and Rui Meng and Yale Song and Xiyu Wei and Sujian Li and Tomas Pfister and Jinsung Yoon},
      year={2026},
      eprint={2601.23265},
      archivePrefix={arXiv},
      primaryClass={cs.CL},
      url={https://arxiv.org/abs/2601.23265}, 
}

@inproceedings{cambricon-sr,
author = {Liu, Tianbo and Song, Xinkai and Yue, Zhifei and Wen, Rui and Hu, Xing and Song, Zhuoran and Wen, Yuanbo and Hao, Yifan and Li, Wei and Du, Zidong and Zhang, Rui and Guo, Jiaming and Huang, Di and Peng, Shaohui and Sun, Guangzhong and Guo, Qi and Chen, Tianshi},
title = {Cambricon-SR: An Accelerator for Neural Scene Representation with Sparse Encoding Table},
year = {2025},
isbn = {9798400712616},
publisher = {Association for Computing Machinery},
address = {New York, NY, USA},
url = {https://doi.org/10.1145/3695053.3731018},
doi = {10.1145/3695053.3731018},
booktitle = {Proceedings of the 52nd Annual International Symposium on Computer Architecture},
pages = {1254–1268},
numpages = {15},
location = {
},
series = {ISCA '25}
}

@inproceedings{F1-paper,
author = {Samardzic, Nikola and Feldmann, Axel and Krastev, Aleksandar and Devadas, Srinivas and Dreslinski, Ronald and Peikert, Christopher and Sanchez, Daniel},
title = {F1: A Fast and Programmable Accelerator for Fully Homomorphic Encryption},
year = {2021},
isbn = {9781450385572},
publisher = {Association for Computing Machinery},
address = {New York, NY, USA},
url = {https://doi.org/10.1145/3466752.3480070},
doi = {10.1145/3466752.3480070},
booktitle = {MICRO-54: 54th Annual IEEE/ACM International Symposium on Microarchitecture},
pages = {238–252},
numpages = {15},
location = {Virtual Event, Greece},
series = {MICRO '21}
}

@inproceedings{williams2009roofline,
  title={Roofline: An Insightful Visual Performance Model for Multicore Architectures},
  author={Williams, Samuel and Waterman, Andrew and Patterson, David},
  booktitle={Communications of the ACM},
  volume={52},
  number={4},
  pages={65--76},
  year={2009},
  publisher={ACM}
}

@article{hill2008amdahl,
  title={Amdahl's law in the multicore era},
  author={Hill, Mark D and Marty, Michael R},
  journal={Computer},
  volume={41},
  number={7},
  pages={33--38},
  year={2008},
  publisher={IEEE}
}

@article{amdahl1967validity,
  title={Validity of the Single Processor Approach to Achieving Large Scale Computing Capabilities},
  author={Amdahl, Gene M.},
  journal={AFIPS Conference Proceedings},
  volume={30},
  pages={483--485},
  year={1967}
}

@inproceedings{ofenbeck2014applying,
  title={Applying the Roofline Model},
  author={Ofenbeck, Georg and Steinmann, Ruedi and Caparros, Victoria and Spampinato, Daniele G. and P{\"u}schel, Markus},
  booktitle={Proceedings of the IEEE International Symposium on Performance Analysis of Systems and Software (ISPASS)},
  pages={76--85},
  year={2014},
  organization={IEEE}
}

@article{gustafson1988reevaluating,
  title={Reevaluating {Amdahl's} Law},
  author={Gustafson, John L.},
  journal={Communications of the ACM},
  volume={31},
  number={5},
  pages={532--533},
  year={1988}
}

@conference{ding2019instruction,
  author       = {Ding, N and Awan, M and Williams, S},
  title        = {Instruction Roofline: An insightful visual performance model for GPUs},
  doi          = {10.1002/cpe.6591},
  url          = {https://www.osti.gov/biblio/1844927},
  place        = {United States},
  organization = {Lawrence Berkeley National Laboratory (LBNL), Berkeley, CA (United States)},
  year         = {2021},
  month        = {01}}

@inproceedings{ivanov2021data,
  title={Data Movement Is All You Need: A Case Study on Optimizing Transformers},
  author={Ivanov, Andrei and Dryden, Nikoli and Ben-Nun, Tal and Li, Shigang and Hoefler, Torsten},
  booktitle={Proceedings of the International Conference on Machine Learning (MLSys)},
  pages={711--722},
  year={2021}
}

@article{kaplan2020scaling,
  title={Scaling Laws for Neural Language Models},
  author={Kaplan, Jared and McCandlish, Sam and Henighan, Tom and Brown, Tom B. and Chess, Benjamin and Child, Rewon and Gray, Scott and Radford, Alec and Wu, Jeffrey and Amodei, Dario},
  journal={arXiv preprint arXiv:2001.08361},
  year={2020}
}

@inproceedings{chowdhery2023palm,
  title={{PaLM}: Scaling Language Modeling with Pathways},
  author={Chowdhery, Aakanksha and Narang, Sharan and Devlin, Jacob and Bosma, Maarten and Mishra, Gaurav and Roberts, Adam and Barham, Paul and Chung, Hyung Won and Sutton, Charles and Gehrmann, Sebastian and others},
  booktitle={Journal of Machine Learning Research},
  volume={24},
  number={240},
  pages={1--113},
  year={2023}
}

@inproceedings{narayanan2021efficient,
  title={Efficient Large-Scale Language Model Training on {GPU} Clusters Using {Megatron-LM}},
  author={Narayanan, Deepak and Shoeybi, Mohammad and Casper, Jared and LeGresley, Patrick and Patwary, Mostofa and Korthikanti, Vijay and Vainbrand, Dmitri and Kasber, Prethvi and Andriasyan, Hayk and Catanzaro, Bryan},
  booktitle={Proceedings of the International Conference for High Performance Computing, Networking, Storage and Analysis (SC)},
  pages={1--15},
  year={2021}
}

@article{binkert2011gem5,
  title={The {gem5} Simulator},
  author={Binkert, Nathan and Beckmann, Bradford and Black, Gabriel and Reinhardt, Steven K. and Saidi, Ali and Basu, Arkaprava and Hestness, Joel and Hower, Derek R. and Krishna, Tushar and Sardashti, Somayeh and others},
  journal={ACM SIGARCH Computer Architecture News},
  volume={39},
  number={2},
  pages={1--7},
  year={2011},
  publisher={ACM}
}

@inproceedings{luo2023ramulator2,
  title={{Ramulator 2.0}: A Modern, Modular, and Extensible {DRAM} Simulator},
  author={Luo, Haocong and Olgun, Ataberk and Yaglikci, A. Giray and Kim, Yoongu and Mutlu, Onur},
  booktitle={IEEE Computer Architecture Letters},
  volume={22},
  number={2},
  pages={101--104},
  year={2023}
}

@article{rodrigues2011sst,
  title={The Structural Simulation Toolkit},
  author={Rodrigues, Arun F. and Hemmert, K. Scott and Barrett, Brian W. and Kersey, Chad and Oldfield, Ron and Weston, Marlo and Risen, Robert and Cook, Jeanine and Rosenfeld, Paul and Cooper-Balis, Elliott and others},
  journal={ACM SIGMETRICS Performance Evaluation Review},
  volume={38},
  number={4},
  pages={37--42},
  year={2011},
  publisher={ACM}
}

@inproceedings{sanchez2013zsim,
  title={{ZSim}: Fast and Accurate Microarchitectural Simulation of Thousand-Core Systems},
  author={Sanchez, Daniel and Kozyrakis, Christos},
  booktitle={Proceedings of the 40th Annual International Symposium on Computer Architecture (ISCA)},
  pages={475--486},
  year={2013},
  organization={ACM}
}

@inproceedings{carlson2011sniper,
  title={Sniper: Exploring the Level of Abstraction for Scalable and Accurate Parallel Multi-Core Simulation},
  author={Carlson, Trevor E. and Heirman, Wim and Eeckhout, Lieven},
  booktitle={Proceedings of the International Conference for High Performance Computing, Networking, Storage and Analysis (SC)},
  pages={1--12},
  year={2011}
}

@inproceedings{lu2024aiscientist,
  title={The {AI} Scientist: Towards Fully Automated Open-Ended Scientific Discovery},
  author={Lu, Chris and Lu, Cong and Lange, Robert Tjarko and Foerster, Jakob and Clune, Jeff and Ha, David},
  booktitle={arXiv preprint arXiv:2408.06292},
  year={2024}
}

@article{liang2024mapping,
  title={Mapping the Increasing Use of {LLMs} in Scientific Papers},
  author={Liang, Weixin and Zhang, Yaohui and Cao, Zhengxuan and Xu, Haifeng and Yu, Yunqi and Schwieterman, Daniel and Kolter, J. Zico and Goldstein, Tom and Hashimoto, Tatsunori},
  journal={arXiv preprint arXiv:2404.01268},
  year={2024}
}

@inproceedings{wadden2020fact,
  title={Fact or Fiction: Verifying Scientific Claims},
  author={Wadden, David and Lin, Shanchuan and Lo, Kyle and Wang, Lucy Lu and van Zuylen, Madelon and Cohan, Arman and Hajishirzi, Hannaneh},
  booktitle={Proceedings of the 2020 Conference on Empirical Methods in Natural Language Processing (EMNLP)},
  pages={7534--7550},
  year={2020}
}

@inproceedings{wright2022citeworth,
  title={CiteWorth: Cite-Worthiness Detection for Improved Scientific Document Understanding},
  author={Wright, Dustin and Augenstein, Isabelle},
  booktitle={Findings of the Association for Computational Linguistics: ACL-IJCNLP 2021},
  pages={1796--1807},
  year={2021}
}

@inproceedings{wu2023autogen,
  title={{AutoGen}: Enabling Next-Gen {LLM} Applications via Multi-Agent Conversation},
  author={Wu, Qingyun and Bansal, Gagan and Zhang, Jieyu and Wu, Yiran and Li, Beibin and Zhu, Erkang and Jiang, Li and Zhang, Xin and Zhang, Shaokun and Liu, Jiale and others},
  booktitle={arXiv preprint arXiv:2308.08155},
  year={2023}
}

@inproceedings{hong2024metagpt,
  title={{MetaGPT}: Meta Programming for a Multi-Agent Collaborative Framework},
  author={Hong, Sirui and Zhuge, Mingchen and Chen, Jonathan and Zheng, Xiawu and Cheng, Yuheng and Wang, Jinlin and Zhang, Ceyao and Wang, Zili and Yau, Steven Ka Shing and Lin, Zijuan and others},
  booktitle={Proceedings of the International Conference on Learning Representations (ICLR)},
  year={2024}
}

@inproceedings{li2023camel,
  title={{CAMEL}: Communicative Agents for ``Mind'' Exploration of Large Language Model Society},
  author={Li, Guohao and Hammoud, Hasan Abed Al Kader and Itani, Hani and Khizbullin, Dmitrii and Ghanem, Bernard},
  booktitle={Proceedings of the 37th Conference on Neural Information Processing Systems (NeurIPS)},
  year={2023}
}

@inproceedings{du2024improving,
  title={Improving Factuality and Reasoning in Language Models through Multiagent Debate},
  author={Du, Yilun and Li, Shuang and Torralba, Antonio and Tenenbaum, Joshua B. and Mordatch, Igor},
  booktitle={Proceedings of the International Conference on Machine Learning (ICML)},
  year={2024}
}

@article{chan2024chateval,
  title={{ChatEval}: Towards Better {LLM}-based Evaluators through Multi-Agent Debate},
  author={Chan, Chi-Min and Chen, Weize and Su, Yusheng and Yu, Jianxuan and Xue, Wei and Zhang, Shanghang and Fu, Jie and Liu, Zhiyuan},
  journal={arXiv preprint arXiv:2308.07201},
  year={2023}
}

@inproceedings{chen2021evaluating,
  title={Evaluating Large Language Models Trained on Code},
  author={Chen, Mark and Tworek, Jerry and Jun, Heewoo and Yuan, Qiming and de Oliveira Pinto, Henrique Pond{\'e} and Kaplan, Jared and Edwards, Harri and Burda, Yuri and Joseph, Nicholas and Brockman, Greg and others},
  booktitle={arXiv preprint arXiv:2107.03374},
  year={2021}
}

@inproceedings{roziere2024code,
  title={Code {Llama}: Open Foundation Models for Code},
  author={Rozi{\`e}re, Baptiste and Gehring, Jonas and Gloeckle, Fabian and Sootla, Sten and Gat, Itai and Tan, Xiaoqing Ellen and Adi, Yossi and Liu, Jingyu and Sauvestre, Romain and Remez, Tal and others},
  booktitle={arXiv preprint arXiv:2308.12950},
  year={2023}
}

@inproceedings{olausson2024selfrepair,
  title={Is Self-Repair a Silver Bullet for Code Generation?},
  author={Olausson, Theo X. and Inala, Jeevana Priya and Wang, Chenglong and Gao, Jianfeng and Solar-Lezama, Armando},
  booktitle={Proceedings of the International Conference on Learning Representations (ICLR)},
  year={2024}
}

@misc{le2024codechain,
      title={CodeChain: Towards Modular Code Generation Through Chain of Self-revisions with Representative Sub-modules}, 
      author={Hung Le and Hailin Chen and Amrita Saha and Akash Gokul and Doyen Sahoo and Shafiq Joty},
      year={2024},
      eprint={2310.08992},
      archivePrefix={arXiv},
      primaryClass={cs.AI},
      url={https://arxiv.org/abs/2310.08992}, 
}

@inproceedings{madaan2023selfrefine,
  title={Self-Refine: Iterative Refinement with Self-Feedback},
  author={Madaan, Aman and Tandon, Niket and Gupta, Prakhar and Hallinan, Skyler and Gao, Luyu and Wiegreffe, Sarah and Alon, Uri and Dziri, Nouha and Prabhumoye, Shravya and Yang, Yiming and others},
  booktitle={Proceedings of the 37th Conference on Neural Information Processing Systems (NeurIPS)},
  year={2023}
}

@inproceedings{shinn2023reflexion,
  title={Reflexion: Language Agents with Verbal Reinforcement Learning},
  author={Shinn, Noah and Cassano, Federico and Gopinath, Ashwin and Narasimhan, Karthik and Yao, Shunyu},
  booktitle={Proceedings of the 37th Conference on Neural Information Processing Systems (NeurIPS)},
  year={2023}
}

@misc{huang2024large,
      title={Large Language Models Cannot Self-Correct Reasoning Yet}, 
      author={Jie Huang and Xinyun Chen and Swaroop Mishra and Huaixiu Steven Zheng and Adams Wei Yu and Xinying Song and Denny Zhou},
      year={2024},
      eprint={2310.01798},
      archivePrefix={arXiv},
      primaryClass={cs.CL},
      url={https://arxiv.org/abs/2310.01798}, 
}

@inproceedings{trinh2024solving,
  title={Solving Olympiad Geometry Without Human Demonstrations},
  author={Trinh, Trieu H. and Wu, Yuhuai and Le, Quoc V. and He, He and Luong, Thang},
  booktitle={Nature},
  volume={625},
  pages={476--482},
  year={2024}
}

@inproceedings{romera2024mathematical,
  title={Mathematical Discoveries from Program Search with Large Language Models},
  author={Romera-Paredes, Bernardino and Barekatain, Mohammadamin and Novikov, Alexander and Balog, Matej and Pranay Kumar, M. and Dupont, Emilien and Ruiz, Francisco J. R. and Ellenberg, Jordan S. and Wang, Pengming and Fawzi, Omar and others},
  booktitle={Nature},
  volume={625},
  pages={468--475},
  year={2024}
}

@article{yang2024leandojo,
  title={{LeanDojo}: Theorem Proving with Retrieval-Augmented Language Models},
  author={Yang, Kaiyu and Swope, Aidan M. and Gu, Alex and Chalamala, Rahul and Song, Peiyang and Yu, Shixing and Godil, Saad and Prenger, Ryan and Anandkumar, Animashree},
  journal={Proceedings of the 37th Conference on Neural Information Processing Systems (NeurIPS)},
  year={2023}
}

@article{polu2022formal,
  title={Formal Mathematics Statement Curriculum Learning},
  author={Polu, Stanislas and Han, Jesse Michael and Zheng, Kunhao and Bousquet-M{\'e}lou, Mireille and Lample, Guillaume and Szegedy, Christian},
  journal={arXiv preprint arXiv:2202.01344},
  year={2022}
}

@article{bai2022constitutional,
  title={Constitutional {AI}: Harmlessness from {AI} Feedback},
  author={Bai, Yuntao and Kadavath, Saurav and Kundu, Sandipan and Askell, Amanda and Kernion, Jackson and Jones, Andy and Chen, Anna and Goldie, Anna and Mirhoseini, Azalia and McKinnon, Cameron and others},
  journal={arXiv preprint arXiv:2212.08073},
  year={2022}
}

@inproceedings{lightman2024lets,
  title={Let's Verify Step by Step},
  author={Lightman, Hunter and Kosaraju, Vineet and Burda, Yuri and Edwards, Harri and Baker, Bowen and Lee, Teddy and Leike, Jan and Schulman, John and Sutskever, Ilya and Cobbe, Karl},
  booktitle={Proceedings of the International Conference on Learning Representations (ICLR)},
  year={2024}
}

@article{cobbe2021training,
  title={Training Verifiers to Solve Math Word Problems},
  author={Cobbe, Karl and Kosaraju, Vineet and Bavarian, Mohammad and Chen, Mark and Jun, Heewoo and Kaiser, Lukasz and Plappert, Matthias and Tworek, Jerry and Hilton, Jacob and Nakano, Reiichiro and others},
  journal={arXiv preprint arXiv:2110.14168},
  year={2021}
}

@misc{brown2024large,
      title={Large Language Monkeys: Scaling Inference Compute with Repeated Sampling}, 
      author={Bradley Brown and Jordan Juravsky and Ryan Ehrlich and Ronald Clark and Quoc V. Le and Christopher Ré and Azalia Mirhoseini},
      year={2024},
      eprint={2407.21787},
      archivePrefix={arXiv},
      primaryClass={cs.LG},
      url={https://arxiv.org/abs/2407.21787}, 
}

@inproceedings{liang2024reviewergpt,
  title={Can Large Language Models Provide Useful Feedback on Research Papers? A Large-Scale Empirical Analysis},
  author={Liang, Weixin and Iber, Yuhui and Li, Yaofei and Cao, Zhengxuan and Chen, Luyu and Hashimoto, Tatsunori and Zou, James},
  booktitle={arXiv preprint arXiv:2310.01783},
  year={2024}
}

@inproceedings{yao2023react,
  title={{ReAct}: Synergizing Reasoning and Acting in Language Models},
  author={Yao, Shunyu and Zhao, Jeffrey and Yu, Dian and Du, Nan and Shafran, Izhak and Narasimhan, Karthik and Cao, Yuan},
  booktitle={Proceedings of the International Conference on Learning Representations (ICLR)},
  year={2023}
}

@inproceedings{wang2024executable,
  title={Executable Code Actions Elicit Better {LLM} Agents},
  author={Wang, Xingyao and Chen, Yangyi and Yuan, Lifan and Zhang, Yizhe and Li, Yunzhu and Peng, Hao and Ji, Heng},
  booktitle={Proceedings of the International Conference on Machine Learning (ICML)},
  year={2024}
}

@inproceedings{dao2022flashattention,
  title={{FlashAttention}: Fast and Memory-Efficient Exact Attention with {IO}-Awareness},
  author={Dao, Tri and Fu, Dan and Ermon, Stefano and Rudra, Atri and R{\'e}, Christopher},
  booktitle={Proceedings of the 36th Conference on Neural Information Processing Systems (NeurIPS)},
  pages={16344--16359},
  year={2022}
}

@article{box1976science,
  title={Science and Statistics},
  author={Box, George E. P.},
  journal={Journal of the American Statistical Association},
  volume={71},
  number={356},
  pages={791--799},
  year={1976}
}

@inproceedings{wei2022chain,
  title={Chain-of-Thought Prompting Elicits Reasoning in Large Language Models},
  author={Wei, Jason and Wang, Xuezhi and Schuurmans, Dale and Bosma, Maarten and Ichter, Brian and Xia, Fei and Chi, Ed and Le, Quoc and Zhou, Denny},
  booktitle={Proceedings of the 36th Conference on Neural Information Processing Systems (NeurIPS)},
  year={2022}
}

@inproceedings{yao2024tree,
  title={Tree of Thoughts: Deliberate Problem Solving with Large Language Models},
  author={Yao, Shunyu and Yu, Dian and Zhao, Jeffrey and Shafran, Izhak and Griffiths, Tom and Cao, Yuan and Narasimhan, Karthik},
  booktitle={Proceedings of the 37th Conference on Neural Information Processing Systems (NeurIPS)},
  year={2024}
}

@inproceedings{lo2020s2orc,
  title={{S2ORC}: The Semantic Scholar Open Research Corpus},
  author={Lo, Kyle and Wang, Lucy Lu and Neumann, Mark and Kinney, Rodney and Weld, Daniel S.},
  booktitle={Proceedings of the 58th Annual Meeting of the Association for Computational Linguistics (ACL)},
  pages={4969--4983},
  year={2020}
}

@article{taylor2022galactica,
  title={Galactica: A Large Language Model for Science},
  author={Taylor, Ross and Kardas, Marcin and Cucurull, Guillem and Scialom, Thomas and Hartshorn, Anthony and Saravia, Elvis and Poulton, Andrew and Kerkez, Viktor and Stojnic, Robert},
  journal={arXiv preprint arXiv:2211.09085},
  year={2022}
}

@inproceedings{ji2023survey,
  title={Survey of Hallucination in Natural Language Generation},
  author={Ji, Ziwei and Lee, Nayeon and Frieske, Rita and Yu, Tiezheng and Su, Dan and Xu, Yan and Ishii, Etsuko and Bang, Ye Jin and Madotto, Andrea and Fung, Pascale},
  booktitle={ACM Computing Surveys},
  volume={55},
  number={12},
  pages={1--38},
  year={2023},
  publisher={ACM}
}

@inproceedings{zheng2023judging,
  title={Judging {LLM}-as-a-Judge with {MT-Bench} and Chatbot Arena},
  author={Zheng, Lianmin and Chiang, Wei-Lin and Sheng, Ying and Zhuang, Siyuan and Wu, Zhanghao and Zhuang, Yonghao and Lin, Zi and Li, Zhuohan and Li, Dacheng and Xing, Eric P. and others},
  booktitle={Proceedings of the 37th Conference on Neural Information Processing Systems (NeurIPS)},
  year={2023}
}

@book{jain1991art,
  title={The Art of Computer Systems Performance Analysis},
  author={Jain, Raj},
  year={1991},
  publisher={John Wiley \& Sons}
}

@book{hennessy2019computer,
  title={Computer Architecture: A Quantitative Approach},
  author={Hennessy, John L. and Patterson, David A.},
  edition={6th},
  year={2019},
  publisher={Morgan Kaufmann}
}

@misc{wang2024autosurvey,
      title={AutoSurvey: Large Language Models Can Automatically Write Surveys}, 
      author={Yidong Wang and Qi Guo and Wenjin Yao and Hongbo Zhang and Xin Zhang and Zhen Wu and Meishan Zhang and Xinyu Dai and Min Zhang and Qingsong Wen and Wei Ye and Shikun Zhang and Yue Zhang},
      year={2024},
      eprint={2406.10252},
      archivePrefix={arXiv},
      primaryClass={cs.IR},
      url={https://arxiv.org/abs/2406.10252}, 
}

@inproceedings{baek2024researchagent,
  title={{ResearchAgent}: Iterative Research Idea Generation over Scientific Literature with Large Language Models},
  author={Baek, Jinheon and Jauhar, Sujay Kumar and Cucerzan, Silviu and Hwang, Sung Ju},
  booktitle={arXiv preprint arXiv:2404.07738},
  year={2024}
}

@inproceedings{li2009mcpat,
  title={{McPAT}: An Integrated Power, Area, and Timing Modeling Framework for Multicore and Manycore Architectures},
  author={Li, Sheng and Ahn, Jung Ho and Strong, Richard D. and Brockman, Jay B. and Tullsen, Dean M. and Jouppi, Norman P.},
  booktitle={Proceedings of the 42nd Annual IEEE/ACM International Symposium on Microarchitecture (MICRO)},
  pages={469--480},
  year={2009}
}

@inproceedings{clearingclouds,
  title={Clearing the Clouds: A Study of Emerging Scale-out Workloads on Modern Hardware},
  author={Ferdman, Michael and Adileh, Almutaz and Ko{\c{c}}berber, Onur and Volos, Stavros and Alisafaee, Mohammad and Jevdjic, Djordje and Kaynak, Cansu and Popescu, Adrian Daniel and Ailamaki, Anastasia and Falsafi, Babak},
  booktitle={Proceedings of the 17th International Conference on Architectural Support for Programming Languages and Operating Systems (ASPLOS)},
  pages={37--48},
  year={2012}
}

@inproceedings{karkhanis2004firstorder,
  title={A First-Order Superscalar Processor Model},
  author={Karkhanis, Tejas S. and Smith, James E.},
  booktitle={Proceedings of the 31st Annual International Symposium on Computer Architecture (ISCA)},
  pages={338--349},
  year={2004},
  organization={IEEE}
}

@article{eyerman2009mechanistic,
  title={A Mechanistic Performance Model for Superscalar Out-of-Order Processors},
  author={Eyerman, Stijn and Eeckhout, Lieven and Karkhanis, Tejas and Smith, James E.},
  journal={ACM Transactions on Computer Systems},
  volume={27},
  number={2},
  pages={1--37},
  year={2009},
  publisher={ACM}
}

@inproceedings{hong2009analytical,
  title={An Analytical Model for a {GPU} Architecture with Memory-Level and Thread-Level Parallelism Awareness},
  author={Hong, Sunpyo and Kim, Hyesoon},
  booktitle={Proceedings of the 36th Annual International Symposium on Computer Architecture (ISCA)},
  pages={152--163},
  year={2009},
  organization={ACM}
}

@inproceedings{huang2014gpumech,
  title={{GPUMech}: {GPU} Performance Modeling Technique Based on Interval Analysis},
  author={Huang, Jen-Cheng and Lee, Joo Hwan and Kim, Hyesoon and Lee, Hsien-Hsin S.},
  booktitle={Proceedings of the 47th Annual IEEE/ACM International Symposium on Microarchitecture (MICRO)},
  pages={268--279},
  year={2014},
  organization={IEEE}
}

@inproceedings{wang2020mdm,
  title={{MDM}: The {GPU} Memory Divergence Model},
  author={Wang, Lu and Jahre, Magnus and Adileh, Almutaz and Eeckhout, Lieven},
  booktitle={Proceedings of the 53rd Annual IEEE/ACM International Symposium on Microarchitecture (MICRO)},
  pages={1089--1101},
  year={2020},
  organization={IEEE}
}

@inproceedings{lee2022gcom,
  title={{GCoM}: A Detailed {GPU} Core Model for Accurate Analytical Modeling of Modern {GPUs}},
  author={Lee, Jounghoo and Ha, Yeonan and Lee, Suhyun and Woo, Jinyoung and Lee, Jinho and Jang, Hanhwi and Kim, Youngsok},
  booktitle={Proceedings of the 49th Annual International Symposium on Computer Architecture (ISCA)},
  pages={424--436},
  year={2022},
  organization={ACM}
}

@inproceedings{sim2012gpuperf,
  title={A Performance Analysis Framework for Identifying Potential Benefits in {GPGPU} Applications},
  author={Sim, Jaewoong and Dasgupta, Aniruddha and Kim, Hyesoon and Vuduc, Richard},
  booktitle={Proceedings of the 17th ACM SIGPLAN Symposium on Principles and Practice of Parallel Programming (PPoPP)},
  pages={11--22},
  year={2012},
  organization={ACM}
}

@inproceedings{baghsorkhi2010adaptive,
  title={An Adaptive Performance Modeling Tool for {GPU} Architectures},
  author={Baghsorkhi, Sara S. and Delahaye, Matthieu and Patel, Sanjay J. and Gropp, William D. and Hwu, Wen-mei W.},
  booktitle={Proceedings of the 15th ACM SIGPLAN Symposium on Principles and Practice of Parallel Programming (PPoPP)},
  pages={105--114},
  year={2010},
  organization={ACM}
}

@inproceedings{zhang2011quantitative,
  title={A Quantitative Performance Analysis Model for {GPU} Architectures},
  author={Zhang, Yao and Owens, John D.},
  booktitle={Proceedings of the 17th IEEE International Symposium on High Performance Computer Architecture (HPCA)},
  pages={382--393},
  year={2011},
  organization={IEEE}
}

@inproceedings{nugteren2014gpu,
  title={A Detailed {GPU} Cache Model Based on Reuse Distance Theory},
  author={Nugteren, Cedric and van den Braak, Gert-Jan and Corporaal, Henk and Bal, Henri},
  booktitle={Proceedings of the 20th IEEE International Symposium on High Performance Computer Architecture (HPCA)},
  pages={37--48},
  year={2014},
  organization={IEEE}
}

@article{kiani2018gpu,
  title={Efficient Cache Performance Modeling in {GPUs} Using Reuse Distance Analysis},
  author={Kiani, Mohsen and Rajabzadeh, Amir},
  journal={ACM Transactions on Architecture and Code Optimization (TACO)},
  volume={15},
  number={4},
  pages={1--24},
  year={2018},
  publisher={ACM}
}

@inproceedings{heo2020realtime,
  title={Real-Time Object Detection System with Multi-Path Neural Networks},
  author={Heo, Seonyeong and Cho, Sungjun and Kim, Youngsok and Kim, Hanjun},
  booktitle={Proceedings of the 26th IEEE Real-Time and Embedded Technology and Applications Symposium (RTAS)},
  pages={174--187},
  year={2020},
  organization={IEEE}
}

@inproceedings{gong2020paqsim,
  title={{PAQSIM}: Fast Performance Model for Graphics Workload on Mobile {GPUs}},
  author={Gong, Xiang and Hu, Chunling and Lim, Chu-Cheow},
  booktitle={Proceedings of the 21st ACM SIGPLAN/SIGBED Conference on Languages, Compilers, and Tools for Embedded Systems (LCTES)},
  pages={127--138},
  year={2020},
  organization={ACM}
}

@inproceedings{villa2021needforspeed,
  title={Need for Speed: Experiences Building a Trustworthy System-Level {GPU} Simulator},
  author={Villa, Oreste and Lustig, Daniel and Yan, Zi and Bolotin, Evgeny and Fu, Yaosheng and Chatterjee, Niladrish and Jiang, Nan and Nellans, David},
  booktitle={Proceedings of the 27th IEEE International Symposium on High-Performance Computer Architecture (HPCA)},
  pages={868--880},
  year={2021},
  organization={IEEE}
}

@inproceedings{arafa2021pptgpu,
  title={Hybrid, Scalable, Trace-Driven Performance Modeling of {GPGPUs}},
  author={Arafa, Yehia and Badawy, Abdel-Hameed and ElWazir, Ammar and Barai, Atanu and Eker, Ali and Chennupati, Gopinath and Santhi, Nandakishore and Eidenbenz, Stephan},
  booktitle={Proceedings of the International Conference for High Performance Computing, Networking, Storage and Analysis (SC)},
  pages={1--15},
  year={2021},
  organization={ACM}
}

@article{TODO-anton,
author = {Shaw, David E. and Deneroff, Martin M. and Dror, Ron O. and Kuskin, Jeffrey S. and Larson, Richard H. and Salmon, John K. and Young, Cliff and Batson, Brannon and Bowers, Kevin J. and Chao, Jack C. and Eastwood, Michael P. and Gagliardo, Joseph and Grossman, J. P. and Ho, C. Richard and Ierardi, Douglas J. and Kolossv\'{a}ry, Istv\'{a}n and Klepeis, John L. and Layman, Timothy and McLeavey, Christine and Moraes, Mark A. and Mueller, Rolf and Priest, Edward C. and Shan, Yibing and Spengler, Jochen and Theobald, Michael and Towles, Brian and Wang, Stanley C.},
title = {Anton, a special-purpose machine for molecular dynamics simulation},
year = {2008},
issue_date = {July 2008},
publisher = {Association for Computing Machinery},
address = {New York, NY, USA},
volume = {51},
number = {7},
issn = {0001-0782},
url = {https://doi.org/10.1145/1364782.1364802},
doi = {10.1145/1364782.1364802},
journal = {Commun. ACM},
month = jul,
pages = {91–97},
numpages = {7}
}

@misc{TODO-basalisc,
      author = {Robin Geelen and Michiel Van Beirendonck and Hilder V. L. Pereira and Brian Huffman and Tynan McAuley and Ben Selfridge and Daniel Wagner and Georgios Dimou and Ingrid Verbauwhede and Frederik Vercauteren and David W. Archer},
      title = {{BASALISC}: Programmable Hardware Accelerator for {BGV} Fully Homomorphic Encryption},
      howpublished = {Cryptology {ePrint} Archive, Paper 2022/657},
      year = {2022},
      doi = {10.46586/tches.v2023.i4.32-57},
      url = {https://eprint.iacr.org/2022/657}
}

@inproceedings{TODO-craterlake,
author = {Samardzic, Nikola and Feldmann, Axel and Krastev, Aleksandar and Manohar, Nathan and Genise, Nicholas and Devadas, Srinivas and Eldefrawy, Karim and Peikert, Chris and Sanchez, Daniel},
title = {CraterLake: a hardware accelerator for efficient unbounded computation on encrypted data},
year = {2022},
isbn = {9781450386104},
publisher = {Association for Computing Machinery},
address = {New York, NY, USA},
url = {https://doi.org/10.1145/3470496.3527393},
doi = {10.1145/3470496.3527393},
booktitle = {Proceedings of the 49th Annual International Symposium on Computer Architecture},
pages = {173–187},
numpages = {15},
location = {New York, New York},
series = {ISCA '22}
}

@inproceedings{TODO-Darwin,
author = {Turakhia, Yatish and Bejerano, Gill and Dally, William J.},
title = {Darwin: A Genomics Co-processor Provides up to 15,000X Acceleration on Long Read Assembly},
year = {2018},
isbn = {9781450349116},
publisher = {Association for Computing Machinery},
address = {New York, NY, USA},
url = {https://doi.org/10.1145/3173162.3173193},
doi = {10.1145/3173162.3173193},
booktitle = {Proceedings of the Twenty-Third International Conference on Architectural Support for Programming Languages and Operating Systems},
pages = {199–213},
numpages = {15},
location = {Williamsburg, VA, USA},
series = {ASPLOS '18}
}

@INPROCEEDINGS{TODO-genax,
  author={Fujiki, Daichi and Subramaniyan, Arun and Zhang, Tianjun and Zeng, Yu and Das, Reetuparna and Blaauw, David and Narayanasamy, Satish},
  booktitle={2018 ACM/IEEE 45th Annual International Symposium on Computer Architecture (ISCA)}, 
  title={GenAx: A Genome Sequencing Accelerator}, 
  year={2018},
  volume={},
  number={},
  pages={69-82},
  doi={10.1109/ISCA.2018.00017}}

@INPROCEEDINGS{TODO-genesis,
  author={Ham, Tae Jun and Bruns-Smith, David and Sweeney, Brendan and Lee, Yejin and Seo, Seong Hoon and Song, U Gyeong and Oh, Young H. and Asanovic, Krste and Lee, Jae W. and Wills, Lisa Wu},
  booktitle={2020 ACM/IEEE 47th Annual International Symposium on Computer Architecture (ISCA)}, 
  title={Genesis: A Hardware Acceleration Framework for Genomic Data Analysis}, 
  year={2020},
  volume={},
  number={},
  pages={254-267},
  doi={10.1109/ISCA45697.2020.00031}}

@inproceedings{TODO-gzkp,
author = {Ma, Weiliang and Xiong, Qian and Shi, Xuanhua and Ma, Xiaosong and Jin, Hai and Kuang, Haozhao and Gao, Mingyu and Zhang, Ye and Shen, Haichen and Hu, Weifang},
title = {GZKP: A GPU Accelerated Zero-Knowledge Proof System},
year = {2023},
isbn = {9781450399166},
publisher = {Association for Computing Machinery},
address = {New York, NY, USA},
url = {https://doi.org/10.1145/3575693.3575711},
doi = {10.1145/3575693.3575711},
booktitle = {Proceedings of the 28th ACM International Conference on Architectural Support for Programming Languages and Operating Systems, Volume 2},
pages = {340–353},
numpages = {14},
location = {Vancouver, BC, Canada},
series = {ASPLOS 2023}
}

@inproceedings{TODO-mdpipe,
author = {Kang, Ning and Yuan, Guojun and Yan, Zihan and Zhang, Beining and Li, Boyang and Li, Zeyu and Wang, Shuo and Chen, Guanglei and Rao, Jiayi and Wang, Zhan and Jia, Weile and Sun, Ninghui and Tan, Guangming},
title = {MD-pipe: A Strong Scaling Enhanced Pipeline Architecture for Ab Initio Accuracy Molecular Dynamics},
year = {2025},
isbn = {9798400712616},
publisher = {Association for Computing Machinery},
address = {New York, NY, USA},
url = {https://doi.org/10.1145/3695053.3731052},
doi = {10.1145/3695053.3731052},
booktitle = {Proceedings of the 52nd Annual International Symposium on Computer Architecture},
pages = {1956–1968},
numpages = {13},
location = {
},
series = {ISCA '25}
}

@inproceedings{TODO-neurex,
author = {Lee, Junseo and Choi, Kwanseok and Lee, Jungi and Lee, Seokwon and Whangbo, Joonho and Sim, Jaewoong},
title = {NeuRex: A Case for Neural Rendering Acceleration},
year = {2023},
isbn = {9798400700958},
publisher = {Association for Computing Machinery},
address = {New York, NY, USA},
url = {https://doi.org/10.1145/3579371.3589056},
doi = {10.1145/3579371.3589056},
booktitle = {Proceedings of the 50th Annual International Symposium on Computer Architecture},
articleno = {21},
numpages = {13},
location = {Orlando, FL, USA},
series = {ISCA '23}
}

@inproceedings{TODO-pipezk,
author = {Zhang, Ye and Wang, Shuo and Zhang, Xian and Dong, Jiangbin and Mao, Xingzhong and Long, Fan and Wang, Cong and Zhou, Dong and Gao, Mingyu and Sun, Guangyu},
title = {PipeZK: accelerating zero-knowledge proof with a pipelined architecture},
year = {2021},
isbn = {9781450390866},
publisher = {IEEE Press},
url = {https://doi.org/10.1109/ISCA52012.2021.00040},
doi = {10.1109/ISCA52012.2021.00040},
booktitle = {Proceedings of the 48th Annual International Symposium on Computer Architecture},
pages = {416–428},
numpages = {13},
location = {Virtual Event, Spain},
series = {ISCA '21}
}

@inproceedings{TODO-segram,
author = {Cali, Damla Senol and Kanellopoulos, Konstantinos and Lindegger, Jo\"{e}l and Bing\"{o}l, Z\"{u}lal and Kalsi, Gurpreet S. and Zuo, Ziyi and Firtina, Can and Cavlak, Meryem Banu and Kim, Jeremie and Ghiasi, Nika Mansouri and Singh, Gagandeep and G\'{o}mez-Luna, Juan and Alserr, Nour Almadhoun and Alser, Mohammed and Subramoney, Sreenivas and Alkan, Can and Ghose, Saugata and Mutlu, Onur},
title = {SeGraM: a universal hardware accelerator for genomic sequence-to-graph and sequence-to-sequence mapping},
year = {2022},
isbn = {9781450386104},
publisher = {Association for Computing Machinery},
address = {New York, NY, USA},
url = {https://doi.org/10.1145/3470496.3527436},
doi = {10.1145/3470496.3527436},
booktitle = {Proceedings of the 49th Annual International Symposium on Computer Architecture},
pages = {638–655},
numpages = {18},
location = {New York, New York},
series = {ISCA '22}
}

@inproceedings{TODO-warehouse,
author = {Ranganathan, Parthasarathy and Stodolsky, Daniel and Calow, Jeff and Dorfman, Jeremy and Guevara, Marisabel and Smullen IV, Clinton Wills and Kuusela, Aki and Balasubramanian, Raghu and Bhatia, Sandeep and Chauhan, Prakash and Cheung, Anna and Chong, In Suk and Dasharathi, Niranjani and Feng, Jia and Fosco, Brian and Foss, Samuel and Gelb, Ben and Gwin, Sara J. and Hase, Yoshiaki and He, Da-ke and Ho, C. Richard and Huffman Jr., Roy W. and Indupalli, Elisha and Jayaram, Indira and Kongetira, Poonacha and Kyaw, Cho Mon and Laursen, Aaron and Li, Yuan and Lou, Fong and Lucke, Kyle A. and Maaninen, JP and Macias, Ramon and Mahony, Maire and Munday, David Alexander and Muroor, Srikanth and Penukonda, Narayana and Perkins-Argueta, Eric and Persaud, Devin and Ramirez, Alex and Rautio, Ville-Mikko and Ripley, Yolanda and Salek, Amir and Sekar, Sathish and Sokolov, Sergey N. and Springer, Rob and Stark, Don and Tan, Mercedes and Wachsler, Mark S. and Walton, Andrew C. and Wickeraad, David A. and Wijaya, Alvin and Wu, Hon Kwan},
title = {Warehouse-scale video acceleration: co-design and deployment in the wild},
year = {2021},
isbn = {9781450383172},
publisher = {Association for Computing Machinery},
address = {New York, NY, USA},
url = {https://doi.org/10.1145/3445814.3446723},
doi = {10.1145/3445814.3446723},
booktitle = {Proceedings of the 26th ACM International Conference on Architectural Support for Programming Languages and Operating Systems},
pages = {600–615},
numpages = {16},
location = {Virtual, USA},
series = {ASPLOS '21}
}


\end{document}